\PassOptionsToPackage{capitalize}{cleveref}

\documentclass[cleveref]{l4dc2027}
\usepackage{preamble}

\extendedpapertrue
\renewcommand{\overviewvariant}{overview-flow}
\newcommand{\extendedversionnote}{This manuscript is an extended version of the paper submitted to L4DC 2027.}

\newif\ifincludeappendix
\includeappendixtrue

\ifincludeappendix\else
  \csundef{r@jmlrstart}
  \csundef{r@jmlrstart@cref}
  \csundef{r@jmlrend}
  \csundef{r@jmlrend@cref}
  \ifextendedpaper
    \csundef{r@fig:tsne-composition}
    \csundef{r@fig:tsne-composition@cref}
    \csundef{r@fig:xland-zero}
    \csundef{r@fig:xland-zero@cref}
  \fi
\fi

\begin{document}

\title[Tensorized Automata]{\papertitle}
\maketitle
\ifextendedpaper
  \ifdefempty{\extendedversionnote}{}{%
    \begingroup
    \renewcommand{\thefootnote}{}%
    \footnotetext{\extendedversionnote}%
    \endgroup}
\fi

\begin{abstract}

Task representations for sequential decision making should describe both an
objective and progress toward it.
We present an algebraically informed pipeline that uses the relational
structure and language semantics of alternating finite automata (AFAs) to
learn control features and compile compact execution operators.
AND-OR message passing learns task and state vectors; selected configurations
define an event-driven Boolean matrix operator.
We prove that initial and successor language coverage make its carrier
represent the exact residual after every prefix, with a sufficient capacity
bound in terms of reachable configurations.
Conjunction and disjunction use the same pipeline.
Fidelity and composition audits evaluate copied formulas and learned disjunct
assignments.
In five-policy-seed XLand experiments with larger training pools, pooled
execution-state vectors outperform raw carrier bits and task-blind
conditioning across three task pools under two rejection rewards.
They achieve 77.3\% success on multi-event tasks and 99.4\% on a four-task
stratum with 33--128-state minimal DFAs, using 16 operator slots.
LetterWorld results show useful residual conditioning alongside a remaining
gap to RAD.
These results connect algebraically informed representation learning to
compact task execution and useful policy conditioning.

\end{abstract}

\begin{keywords}
  Learning for control, task-conditioned reinforcement learning, temporal task specifications,
  representation learning, alternating finite automata
\end{keywords}

\section{Introduction}
\label{sec:intro}

Task-conditioned policies share representations across tasks
\citep{schaul2015universal,hansen-estruch2022bisimulation}. Temporal tasks
require both an objective and its progress: ``visit \(a\) and \(b\), in
either order, then \(c\)'' combines concurrent obligations with sequence.
Automata and reward machines provide explicit event-language semantics
\citep{icarte2022reward,camacho2019ltl}. We seek representations supporting
faithful execution and useful progress features for shared control.

Tasks are represented as alternating finite automata (AFAs). Their transitions
combine disjunctive alternatives and conjunctive obligations, providing direct
constructions for composition. AFAs can be exponentially more succinct than
NFAs and doubly exponentially more succinct than DFAs
\citep{chandra1981alternation,dantoni2018symbolic}, and support translations
from temporal logic \citep{vardi1995alternating}. These structures describe
regular languages, with multiple automata potentially describing the same
language. We use this semantics to supervise learned representations across
presentations.

Our design uses the relational structure of AFAs and their conjunction,
disjunction, and residual semantics throughout learning and execution.
AND-OR hypergraph message passing learns task and state vectors;
configuration selection and semantic compilation produce a compact Boolean
matrix operator. Its active configurations form an event-driven carrier,
whose learned state vectors provide features for a shared policy. We study
copied formulas and learned disjunct assignments, retaining input conjuncts
and acceptance data, and characterize when the carrier represents the exact
remaining language.

Conjunctive obligations are carried within configuration slots, while active
slots express alternatives. After compilation, Boolean matrix updates track
progress and expose learned control features without online symbolic formula
manipulation or residual re-encoding. This separates the execution interface
from explicit DFA construction: a compact carrier can represent tasks whose
minimal DFAs have many more states than the operator has slots.

\ifextendedpaper
\begin{figure}[t]
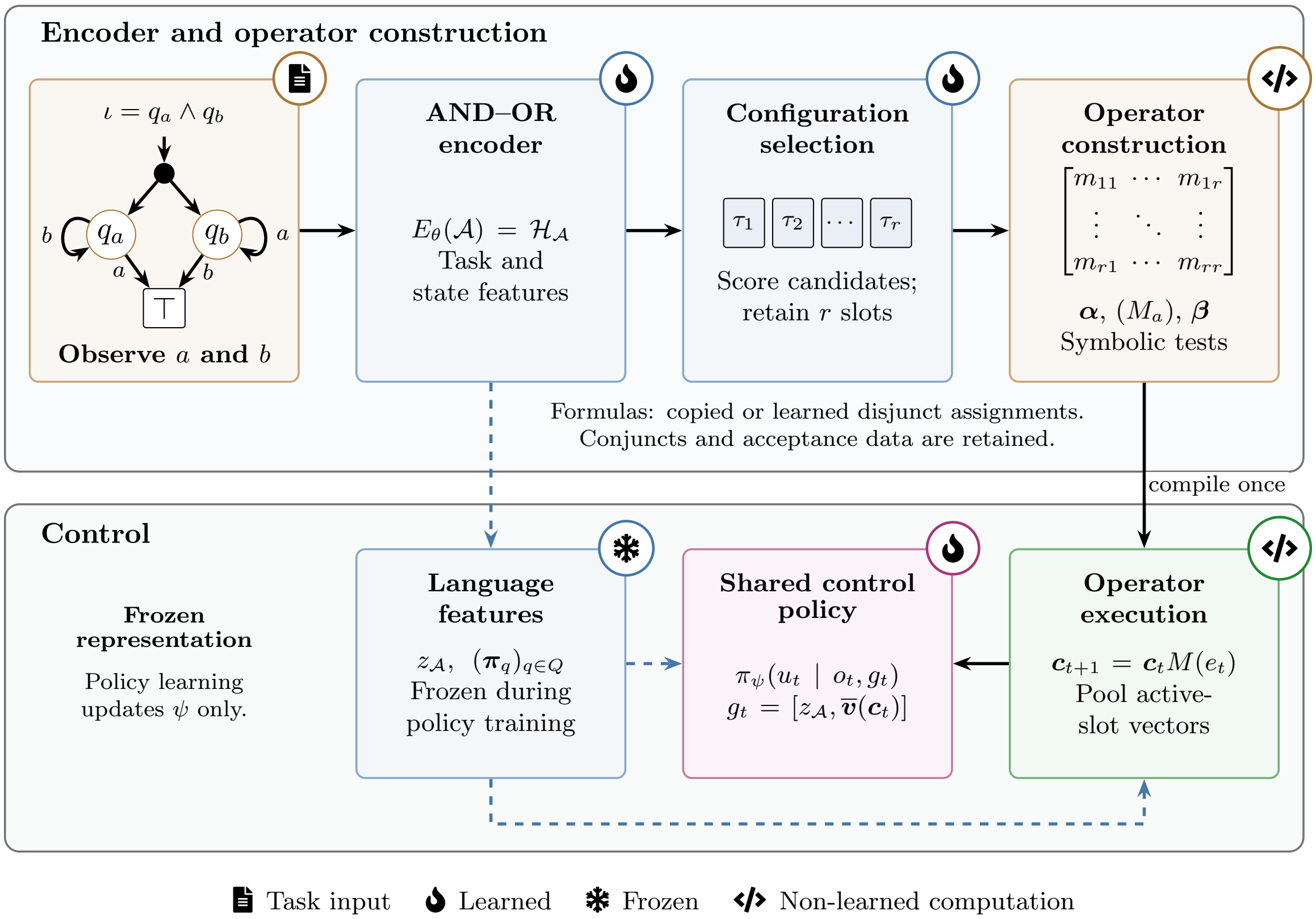

  \centering
  \includestandalone[width=\linewidth]{figures/standalone/\overviewvariant}
  \caption{Algebraically informed task representation and control. The encoder
    learns task and state features; learned scores select configuration slots.
    Formula sources are copied or use learned disjunct assignments over copied
    conjuncts. Symbolic tests construct the operator, and Boolean matrix computations track
    event-driven progress. Active-slot vectors and the initial task vector
    condition a separately trained policy. Exact residual execution requires
    initial and successor language coverage.}
  \label{fig:overview}
  \vspace{-1.5\baselineskip}
\end{figure}
\fi

Our contributions are:
(i) a structured AFA representation with algebraic message passing and a
compiled operator, supporting conjunction and disjunction through the same
construction;
(ii) an exact residual-execution theorem under language-coverage conditions,
with a sufficient configuration-capacity bound; and
(iii) empirical evidence that pooled execution-state vectors provide useful
control features, including on tasks with minimal DFAs larger than the
operator budget, supported by fidelity and composition audits.

\subsection{Related Work}
\label{sec:related}

\mypara{Representations for temporal tasks.}
LTL2Action encodes temporal-logic instructions and progresses their formulas
\citep{vaezipoor2021ltl2action}. RAD encodes conjunctive DFA compositions
through a graph attention network, with a special node for conjunction and
pretraining on reach-avoid-derived tasks \citep{yalcinkaya2024compositional}.
Its later bisimulation-metric objective studies guarantees for optimal DFA
encoders \citep{yalcinkaya2025provably}. Our AFA representation uses
conjunction and disjunction within both initial and transition formulas.

\mypara{Algebraic neural computation.}
Neural Bellman-Ford Networks parameterize path computations with learned
messages and aggregation, including a max-min formulation \citep{zhu2021neural}.
Neural theorem proving combines minima along a proof with maxima over
alternative proofs \citep{rocktaschel2017endtoend}. Learnable aggregation
\citep{li2023deepergcn}, max/min algorithm learning \citep{wittig2026which},
and fuzzy graph aggregation \citep{xie2026fgmma} provide related precedents.

\mypara{Neural automata and matrix operators.}
\citet{dhayalkar2026logicgated} constructs or trains a logic-gated network per
target AFA and evaluates word acceptance; our encoder and configuration
selector share parameters across task automata. The operator builds on
differentiable automata
\citep{balakrishnan2024motion,balakrishnan2024differentiable} and weighted-AFA
monitoring \citep{balakrishnan2025monitoring}, with reachable carriers forming
finite-state reward machines \citep{icarte2022reward}.

\section{Preliminaries}
\label{sec:prelim}

Let \(2^X\) denote the powerset of \(X\), and \(\Sigma^*\) the finite words
over alphabet \(\Sigma\), including \(\varepsilon\).
For words \(u,v\), \(uv\) denotes concatenation and \(\abs u\) length.
Let \(\PBool[Q]\) contain positive Boolean formulas over states \(Q\), built
using conjunction, disjunction, and constants \(\top,\bot\).

\begin{definition}[Alternating finite automaton]
  An AFA is \(\Ac=\Tuple{\Sigma,Q,\iota,\Delta,F}\), with states \(Q\),
  initial formula \(\iota\in\PBool[Q]\), transition map
  \(\Delta:Q\times\Sigma\to\PBool[Q]\), and accepting states \(F \subseteq
  Q\).
\end{definition}
For \(a\in\Sigma\), we overload notation by extending transitions to formulas
\(p \in \PBool[Q]\) by substitution and composition,
\begin{equation}
  \begin{array}
    {c@{\qquad}c@{\qquad}c}
    \Delta_a(p)=p[q\leftarrow\Delta(q,a)]_{q\in Q}, &
    \Delta_\varepsilon(p)=p,                        &
    \Delta_{ua}(p)=\Delta_a(\Delta_u(p)).
  \end{array}
  \label{eq:formula-step}
\end{equation}
A word \(w\) is \emph{accepted} when \(\Delta_w(\iota)\) evaluates to true
under the assignment \(q\mapsto\mathbf{1}[q\in F]\).

\(L(\Ac) \subset \Sigma^*\) denotes the \emph{language} accepted by the
automaton \(\Ac\), and we let \(L_\Ac(p)\) denote the language obtained by
using \(p \in \PBool[Q]\) as the initial formula for \(\Ac\).
More generally, for some \(L\subseteq\Sigma^*\) and \(u\in\Sigma^*\), we
define the \emph{residual language} \(u^{-1}L=\Set{v\in\Sigma^*\given uv\in
  L}\) \citep{brzozowski1964derivatives}.
Let \(\Ac_u\) denote the automaton obtained by replacing \(\iota\) with
\(\Delta_u(\iota)\) in \(\Ac\), such that \(L(\Ac_u)=u^{-1}L(\Ac)\).

For AFAs \(\Ac_1,\Ac_2\) over the same alphabet with disjoint state sets,
conjunction and disjunction use state set \(Q_1\sqcup Q_2\), transition map
\(\Delta_1\sqcup\Delta_2\), and accepting set \(F_1\sqcup F_2\).
Initial formula \(\iota_1\land\iota_2\) gives language \(L(\Ac_1)\cap
L(\Ac_2)\); initial formula \(\iota_1\lor\iota_2\) gives \(L(\Ac_1)\cup
L(\Ac_2)\).

\begin{definition}[Configuration]
  A \emph{configuration} of \(\Ac\) is a set \(\tau\subseteq Q\), which we
  interpret as the simultaneous obligations \(\bigwedge_{q\in\tau}q\), with
  language \(L_\Ac(\tau)=\bigcap_{q\in\tau}L_\Ac(q)\).
  We let \([p]\subseteq 2^Q\) denote the configurations appearing as terms in
  the irredundant DNF of \(p\in\PBool[Q]\).
  In particular, \([\top]=\{\emptyset\}\), \([\bot]=\emptyset\), and
  \(L_\Ac(\emptyset)=\Sigma^*\).
\end{definition}

\emph{Dealternation} converts an AFA into an equivalent configuration NFA
that has configurations as states, the terms of \(\iota\) as initial states,
\(\tau\subseteq F\) as accepting states, and transitions
\begin{equation}
  \tau\xrightarrow{a}\tau'
  \quad\Longleftrightarrow\quad
  \tau'\in\left[\bigwedge_{q\in\tau}\Delta(q,a)\right].
  \label{eq:configuration-step}
\end{equation}
The NFA's minimal DFA \(D_\Ac\) has at most \(2^C\) states.
A configuration records simultaneous obligations and an active set of terms records
alternatives.
Letting \({\to}=\bigcup_{a\in\Sigma}{\xrightarrow{a}}\), the reflexive
transitive closure \(\to^*\) can be used to define the reachable
configurations of the AFA
\begin{equation}
  \Reachable(\Ac)=\Set{\tau\subseteq Q\given
    \exists\tau_0\in[\iota],\ \tau_0\to^*\tau},\qquad
  C=\abs{\Reachable(\Ac)}\leq2^{\abs Q}.
\end{equation}

More generally, we can represent an NFA as a \emph{matrix operator}.
\begin{definition}[Matrix Operator]
  The matrix operator of an \(r\)-state NFA is a tuple
  \(\left(\vect{\alpha}, \vect{\beta}, \Set*{M_a}_{a \in \Sigma}\right)\),
  where \(\vect{\alpha},\vect{\beta} \in \Be^r\) are initial and accepting
  vectors, and \(M_a \in \Be^{r \times r}\) are the transition matrices such
  that
  \(M_a[i,j]=\mathbf{1}[\tau_i\xrightarrow{a}\tau_j]\),
  \(\alpha_i=\mathbf{1}[\tau_i\in[\iota]]\), and
  \(\beta_i=\mathbf{1}[\tau_i\subseteq F]\).
  The Boolean row vector \(\vect{c}_u\), called the carrier, records active states:
  \begin{equation}
    \vect{c}_\varepsilon=\vect{\alpha}^{\mathsf T},\qquad
    \vect{c}_{ua}=\vect{c}_uM_a,\qquad
    f_\Ac(w)=\vect{c}_w\vect{\beta}.
    \label{eq:carrier}
  \end{equation}
  Acceptance holds exactly when \(f_\Ac(w)=1\).
\end{definition}
For a word
\(w=a_1\cdots a_k\), write \(M(w)=M_{a_1}\cdots M_{a_k}\),
with \(M(\varepsilon)=I_r\).
In the sequel, we will omit the task subscript when clear.
Also, note that for Boolean matrices \(A\in\Be^{m\times n}\) and
\(B\in\Be^{n\times p}\), their product \(AB\in\Be^{m\times p}\) is
\((AB)_{ij}=\bigvee_{k=1}^{n}(A_{ik}\land B_{kj})\).

\section{Algebraic Task Representations}
\label{sec:representation}

The pipeline uses AFA semantics at each stage: conjunction and disjunction
guide message aggregation, configuration languages give operator states
their meaning, and language membership and equivalence supply supervision.
Learned task and state vectors accompany a compiled execution operator:
\begin{equation}
  \Ac\xmapsto{E_\theta}\mathcal{H}_\Ac
  \xmapsto{D_\eta}
  \Tuple{\vect{\alpha}_\Ac,(M_{\Ac,a})_{a\in\Sigma},\vect{\beta}_\Ac}.
  \label{eq:representation-pipeline}
\end{equation}
The encoder \(E_\theta\) and decoder \(D_\eta\) share parameters across
automata; the decoded NFA has \(r\) states, called \emph{slots}.

\subsection{Hypergraph encoding and algebraic message passing}

We define the \emph{hypergraph} encoding of an AFA with the vertices \(V= Q
\sqcup \Set{v_\iota} \) (where \(v_\iota\) is a new node corresponding
to \(\iota\)) and hyperedge set
\begin{equation}
  E = \Set[\Big]{(q,\tau,a)\given q\in Q,a\in\Sigma,\tau\in[\Delta(q,a)]}
  \sqcup \Set[\Big]{(v_\iota,\tau,\mathrm{init})\given\tau\in[\iota]}.
  \label{eq:hypergraph}
\end{equation}
An edge \(e=(v,\tau,a)\) has source \(v\) and target set \(\tau\).
Let \(m = \abs{\Sigma}\), then with \(\vect e_j\in\Be^m\) the \(j\)-th basis
vector, assign vertex and edge features corresponding to initial states,
accepting states, and transition guards as
\begin{equation}
  \begin{aligned}
    x_v & = \bigl(\mathbf1[v\in F],\mathbf1[v=v_\iota]\bigr)\in\Be^2,
        &                                                             & v\in V,          \\
    x_e & =
    \begin{cases}
      (\vect e_j,0), & e=(q,\tau,a_j),                 \\
      (\vect0_m,1),  & e=(v_\iota,\tau,\mathrm{init}),
    \end{cases}
        &                                                             & x_e\in\Be^{m+1}.
  \end{aligned}
  \label{eq:hypergraph-features}
\end{equation}

\mypara{Algebraic message passing.}
Let \(h_v^{(\ell)}\in\Re^d\) denote the vector for vertex \(v\) after
layer \(\ell\), \(T_e\) the targets of edge \(e\), and \(S_{q,a}\) the
hyperedges from \(q\) labeled by \(a\).
Coordinatewise smooth minimum
and maximum, with learned empty-pool vectors, aggregate targets and alternatives:
\begin{equation}
  \begin{array}{c}
    k_e^{(\ell)} =\smin_{\epsilon_\ell}\{h_p^{(\ell)}:p\in T_e\},\qquad
    p_{q,a}^{(\ell)} =\smax_{\epsilon_\ell}\{k_e^{(\ell)}:e\in S_{q,a}\},
    \\[2ex]
    h_q^{(\ell+1)} =\operatorname{LayerNorm}\!
    \left(h_q^{(\ell)}+ U_\ell\!\left([\mathbf{1}[q\in F],(p_{q,a}^{(\ell)})_{a\in\Sigma}]\right)\right).
  \end{array}
\end{equation}
The initial vertex uses updated state vectors and its own hyperedges and
residual map, consuming no input symbol.
The maps \(U_\ell\) and positive
temperatures \(\epsilon_\ell\) are learned.
On Boolean membership values,
the zero-temperature pools implement the one-symbol language recursion;
learned hidden vectors additionally pass through residual maps.

\mypara{Structured encoder output.}
After \(L\) layers, we read the task vector from the initial vertex and
retain a vector for each source state:
\begin{equation}
  z_\Ac=h_{v_\iota}^{(L)}\in\Re^d,\qquad
  h_q=h_q^{(L)},\qquad
  \vect{\pi}_q=\operatorname{LayerNorm}(W_Vh_q)\in\Re^{d_\pi}.
  \label{eq:task-latent}
\end{equation}
The structured output \(\mathcal H_\Ac\) contains \(z_\Ac\), the
source-aligned vectors \(h_q,\vect\pi_q\), formula factors, and input
acceptance data.
The decoder consumes this full output.

\subsection{Constructing the operator}

We consider two pipeline regimes: one copies the input formulas, while the
other learns disjunct assignments over copied conjuncts.
Both retain input
acceptance data and use configuration selection to construct the execution
operator.

\mypara{Formula sources.}
Copied formulas use \((\widehat\iota,\widehat\Delta)=(\iota,\Delta)\).
For learned assignments, index a formula by \(g\in(Q\times\Sigma)\sqcup\{\iota\}\).
Its eligible edges are \(E_{(q,a)}=S_{q,a}\) for transitions and
\(E_\iota=\{(v_\iota,\tau,\mathrm{init}):\tau\in[\iota]\}\) initially.
For each of \(R\) disjunct positions, the head predicts weights
\(w_{gje}=w_{gje}(\mathcal H_\Ac;\eta)\geq0\), with \(\sum_{e\in E_g\sqcup\{\bot\}}w_{gje}=1\);
the extra choice \(\bot\) suppresses a disjunct.

Let \(c_{ek}\in Q\sqcup\{\top\}\) be the \(k\)-th target of edge \(e\)
in source-state order, padded with \(\top\) to \(K\) conjunct positions.
The copied conjunct indicators are mixed by these weights:
\begin{equation}
  P_{gjk}(x)=\sum_{e\in E_g}w_{gje}\mathbf1[c_{ek}=x]
  +w_{gj\bot}\mathbf1[x=\bot],\qquad
  x\in Q\sqcup\{\top,\bot\}.
  \label{eq:assignment-factors}
\end{equation}
Thresholding gives the assigned terms and formulas:
\begin{equation}
  \begin{array}{c}
    \widehat\tau_{gj}
    =\Set[\Big]{q\in Q\given\max_{1\leq k\leq K}P_{gjk}(q)>1/2}
    \qquad
    \displaystyle\widehat\psi_g
    =\bigvee_{\substack{1\leq j\leq R   \\
    w_{gj\bot}<1/2}}
    \bigwedge_{q\in\widehat\tau_{gj}}q, \\[.75ex]
    \widehat\iota=\widehat\psi_\iota
    \qquad
    \widehat\Delta(q,a)=\widehat\psi_{(q,a)}.
  \end{array}
  \label{eq:formula-sources}
\end{equation}
This defines the disjunct-assignment map \(\mathcal F_\eta\).
Initial and transition formulas have separate learned weights.
\cref{app:formula-selection} gives their normalization rules and caps.
Both regimes retain the accepting set \(F\).

\mypara{Configuration selection.}
Let \(\mathcal C_N\subseteq2^Q\) contain the configurations obtained by
breadth-first exploration from \([\widehat\iota]\), expanding each \(\tau\)
through \([\bigwedge_{q\in\tau}\widehat\Delta(q,a)]\) for each
\(a\in\Sigma\), until closure or the budget \(N\).
A shared score \(s_\eta(\tau;\mathcal H_\Ac)\)
selects the live slots:
\begin{equation}
  k=\min(r,\abs{\mathcal C_N}),\qquad
  (\tau_1,\ldots,\tau_k)
  =\operatorname{Top}_{k}\!
  \left(\mathcal C_N;
  s_\eta(\cdot;\mathcal H_\Ac)\right).
  \label{eq:configuration-selection}
\end{equation}
All candidates are retained when they fit; the remaining \(r-k\) slots are
inert padding.
For copied formulas, completed exploration gives
\(\mathcal C_N=\Reachable(\Ac)\).
Learned assignments also determine which
configurations enter \(\mathcal C_N\).
\cref{app:formula-selection} specifies the scoring network and gradient estimator.

\mypara{Operator construction.}
Each live slot represents the input language \(\ell_t=L_\Ac(\tau_t)\).
Syntactic construction tests the assigned formulas; semantic construction
tests language inclusion in the input AFA.
For live indices \(t,t'\),
\begin{equation}
  \begin{aligned}
    \alpha_t^{\mathrm{syn}}
            & =\mathbf1[\tau_t\models\widehat\iota],
            & \alpha_t^{\mathrm{sem}}
            & =\mathbf1[\ell_t\subseteq L(\Ac)],          \\
    M_a^{\mathrm{syn}}[t,t']
            & =\mathbf1\!
    \left[\tau_{t'}\models
      \bigwedge_{q\in\tau_t}\widehat\Delta(q,a)\right],
            & M_a^{\mathrm{sem}}[t,t']
            & =\mathbf1[\ell_{t'}\subseteq a^{-1}\ell_t], \\
    \beta_t & =\mathbf1[\tau_t\subseteq F]
    =\mathbf1[\varepsilon\in\ell_t].
  \end{aligned}
  \label{eq:semantic-operator}
\end{equation}
Padded entries are zero.
Semantic export uses exact language inclusion;
training uses the surrogate described below.
The resulting Boolean arrays
execute through \eqref{eq:carrier}.

\subsection{Semantic supervision}
\label{sec:training}

Pretraining uses automata, their symbolic residuals, and words with exact
acceptance labels. Let \(s_\Ac(w)=\widetilde f_\Ac(w)\in[0,1]\) be the
soft operator's word score, computed by max-min multiplication, and let
\(y_\Ac(w)=\mathbf{1}[w\in L(\Ac)]\). We combine six objectives:
\begin{equation}
 \Lc=\lambda_{\mathrm{mem}}\Lc_{\mathrm{mem}}
 +\lambda_{\mathrm{pre}}\Lc_{\mathrm{pre}}
 +\lambda_{\mathrm{wit}}\Lc_{\mathrm{wit}}
 +\lambda_{\mathrm{vertex}}\Lc_{\mathrm{vertex}}
 +\lambda_{\mathrm{id}}\Lc_{\mathrm{id}}
 +\lambda_{\mathrm{tr}}\Lc_{\mathrm{tr}}.
 \label{eq:training-objective}
\end{equation}
Averages use valid sampled entries. For threshold \(b\), margin \(m\geq0\),
and \([x]_+=\max(x,0)\), define the squared hinge
\(\ell_{b,m}(s,y)=y[b+m-s]_+^2+(1-y)[s-b+m]_+^2\).

\mypara{Membership and prefixes.}
The membership set \(\mathcal D_{\mathrm{mem}}\) contains labeled words
for input and sampled residual automata. The prefix set
\(\mathcal D_{\mathrm{pre}}\) contains every prefix, including
\(\varepsilon\), of sampled long words, labeled on its source automaton.
Both sets use the same acceptance penalty:
\begin{equation}
 \Lc_k=\operatorname{BalAvg}_{(\Ac,w)\in\mathcal D_k}
 \ell_{b,m_{\mathrm{mem}}}\bigl(s_\Ac(w),y_\Ac(w)\bigr),
 \qquad k\in\{\mathrm{mem},\mathrm{pre}\}.
 \label{eq:membership-prefix-objectives}
\end{equation}
Here \(\operatorname{BalAvg}\) balances labels and length buckets;
\cref{app:sampling,app:objectives} specify averaging and sampling conventions.

\mypara{Distinguishing witnesses.}
For \(w\in L(\Ac^+)\setminus L(\Ac^-)\), let
\(s^+=s_{\Ac^+}(w)\) and \(s^-=s_{\Ac^-}(w)\). For margin \(m_{\mathrm{wit}}\geq0\), the witness objective
\(\Lc_{\mathrm{wit}}\) averages
\begin{equation}
 \ell_{\mathrm{wit}}(s^+,s^-)=
 \ell_{b,m_{\mathrm{mem}}}(s^+,1)+\ell_{b,m_{\mathrm{mem}}}(s^-,0)
 +[m_{\mathrm{wit}}-(s^+-s^-)]_+^2.
 \label{eq:witness-loss}
\end{equation}

\mypara{Vertex languages.}
A linear head on \(h_v^{(L)}\) predicts logits \(a_{v,w}\) for words in
\(\mathcal W\). With \(L_v=L_\Ac(v)\) at source states and
\(L_{v_\iota}=L(\Ac)\), the loss is
\[
 \Lc_{\mathrm{vertex}}=\operatorname{Avg}_{v,w}
 \operatorname{BCEWithLogits}(a_{v,w},\mathbf1[w\in L_v]),
\]
with the average over valid vertices and table words.

\mypara{Task-vector identity.}
For task vectors \(z,z'\), languages \(L,L'\), and shortest
distinguishing-word length \(j\) when \(L\ne L'\), define
\begin{equation}
 \ell_{\mathrm{id}}(z,z')=
 \begin{cases}
 \norm{z-z'}_2^2,&L=L',\\
 [m_{\mathrm{id}}\rho^j-\norm{z-z'}_2]_+^2,&L\ne L',
 \end{cases}\qquad 0<\rho\leq1.
 \label{eq:identity-loss}
\end{equation}
The objective \(\Lc_{\mathrm{id}}\) adds the mean over equal-language
pairs and the mean over different-language pairs.

\mypara{Formula probes.}
For an input formula \(\phi\in\{\iota\}\cup
\{\Delta(q,a):q\in Q,a\in\Sigma\}\), a probe \(S\subseteq Q\)
has target \(y_\phi(S)=\mathbf{1}[S\models\phi]\), where states in
\(S\) are true. Its decoded soft evaluation \(\widetilde\phi(S)\)
receives penalty \(\ell_{.5,.2}(\widetilde\phi(S),y_\phi(S))\).
The objective \(\Lc_{\mathrm{tr}}\) sums label-balanced probe penalties
for transition and initial formulas and acceptance targets \(\mathbf1[q\in F]\).
Copied formulas and acceptance bits satisfy their targets exactly.

Operator losses supervise candidate selection through its effect on
acceptance. During training, bounded word tables, witnesses, and separators
approximate language inclusion with a differentiable surrogate. At export,
we use exact inclusion in the input AFA to construct the Boolean arrays.
\cref{app:training,app:surrogate} give weights and surrogate limitations.

\section{Task Execution and Sequential Decision Making}
\label{sec:properties}
\label{sec:decision-making}

\begin{figure}[t]
  \centering
  \includestandalone[]{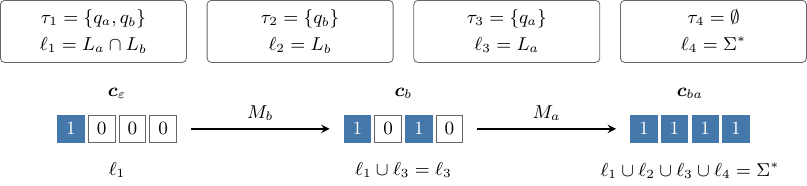}
  \caption{Semantic execution of ``observe both $a$ and $b$, in any order.''
    For $\Sigma=\{a,b\}$, state $q_x$ requires event $x$, with language
    $L_x=\Sigma^*x\Sigma^*$: $\Delta(q_x,x)=\top$, and other events leave
    $q_x$ unchanged. Initially $\iota=q_a\wedge q_b$, with $F=\emptyset$.
    Top: the four retained configurations and their languages.
    Bottom: Boolean carrier updates along $ba$; filled entries mark active
    slots in order $1$--$4$. Their language union, shown below each carrier,
    is the exact residual. After $b$, $\ell_1\subseteq\ell_3$, so the extra
    active slot preserves the residual; after $ba$, slot $4$ ensures acceptance.
    \cref{app:running-example} gives the full construction and policy features.}
  \label{fig:execution-example}
\end{figure}

Each active slot represents a configuration language; their union describes
remaining obligations. \cref{fig:execution-example} illustrates how several
active slots can represent the same residual as a single slot. The following
theorem gives conditions for exact residual execution.

\begin{theorem}[Exact residual execution]
  \label{thm:residual-execution}
  Let \(\Ac\) be an AFA and let
  \((\vect{\alpha},(M_a)_{a\in\Sigma},\vect{\beta})\) be its semantic
  operator constructed by \eqref{eq:semantic-operator}, with live slots
  \(J\) representing languages \(\ell_t=L_\Ac(\tau_t)\) and padded
  entries zero.
  If these languages satisfy initial and successor coverage,
  the carrier tracks the residual language of \(\Ac\) exactly.
  Specifically, if
  \begin{equation}
    \begin{array}{c@{\quad\textnormal{and}\quad}c}
      L(\Ac)=\bigcup_{\substack{t\in J          \\
      \ell_t\subseteq L(\Ac)}}\ell_t,
      \label{eq:initial-coverage}
       &
      \forall t \in J, a \in \Sigma, \quad
      a^{-1}\ell_t = \bigcup_{\substack{t'\in J \\
      \ell_{t'}\subseteq a^{-1}\ell_t}}\ell_{t'}
    \end{array}
  \end{equation}
  then, for every word \(u\in\Sigma^*\),
  \begin{equation}
    u^{-1}L(\Ac)=\bigcup_{\substack{t\in J\\
    c_u[t]=1}}\ell_t.
    \label{eq:residual-operator}
  \end{equation}
\end{theorem}

Induction on the event word gives \eqref{eq:residual-operator}; the full
proof appears in \cref{app:residual-proof}.
Acceptance follows immediately:
\(\vect c_u\vect\beta=1\) if and only if \(u\in L(\Ac)\).

\begin{corollary}[Sufficient configuration capacity]
  Selecting all \(C\) reachable configurations satisfies the coverage
  conditions of \cref{thm:residual-execution}.
  Consequently,
  \(C\leq r\) suffices for an exact \(r\)-slot semantic operator.
  The syntactic construction is also exact when it uses the input formulas
  and the same complete configuration family.
  \label{cor:capacity}
\end{corollary}

\begin{remark}
  Each slot represents a conjunction of source-state obligations; active slots
  express alternatives. The complete \(C\)-configuration construction admits
  at most \(2^C\) carriers, giving \(\abs{D_\Ac}\leq2^C\), with distinct
  DFA states corresponding to reachable residual languages. Thus exact
  execution can use fewer slots than DFA states; any smaller configuration
  family preserving coverage also suffices. This guarantee concerns execution;
  control-feature quality is evaluated empirically.
\end{remark}

\mypara{Policy conditioning.}
The carrier identifies active configuration languages; pooling their learned
state vectors exposes this execution state to a shared policy
\(\pi_\psi(u_t\mid o_t,g_t)\), where \(o_t\) is the physical observation,
\(u_t\) is the control action, and \(g_t\) is the conditioning vector.
A labeling function maps physical transitions to event words \(e_t\in\Sigma^*\),
giving prefix \(w_t=e_0\cdots e_{t-1}\). Four conditioning maps are:
\begin{center}
\begin{tabular}{lcccc}
\toprule
Input & Initial & Residual & Carrier bits & Carrier vectors\\
\midrule
\(g_t\) & \(z_\Ac\) & \(z_{\Ac_{w_t}}\) &
\([z_\Ac,\vect{c}_{w_t}]\) &
\([z_\Ac,\overline{\vect{v}}(\vect{c}_{w_t})]\)\\
\bottomrule
\end{tabular}
\end{center}
For the vector interface, we give each configuration slot \(j\) the mean
of its pinned source-state vectors and pool the active slots:
\begin{equation}
  \vect{v}_j=
    \frac{\sum_{q\in\tau_j}\vect{\pi}_q}{\max(1,\abs{\tau_j})},\qquad
  \overline{\vect{v}}(\vect{c})=
    \frac{\sum_j c[j]\vect{v}_j}{\max(1,\sum_j c[j])}.
  \label{eq:carrier-vector}
\end{equation}
Both carrier interfaces keep \(z_\Ac\) fixed and update
\(\vect c_{w_{t+1}}=\vect c_{w_t}M(e_t)\). Carrier bits index task-specific
slots; \(\overline{\vect v}\) averages their active configuration vectors.
Empty configurations and empty carriers yield zero vectors.

\section{Experimental Evaluation}
\label{sec:experiments}

We evaluate execution by bounded-word agreement, composition by all-word
equivalence, and held-out control by success and matched interface differences.\footnote{Source code for the experiments will be made available at
  \url{https://git.anandb.dev/morphata.git}, with documentation at
  \url{https://docs.anandb.dev/morphata/}.}

\subsection{Experimental setup}

\mypara{Representation learning.}
Pretraining uses automata with 2--8 source states and an eight-layer AND-OR
encoder of width 32.
Source-state and operator capacities are 16, with at
most 64 configuration candidates.
Copied formulas use syntactic transitions;
learned assignments use semantic transitions and identity supervision.
\cref{app:training} gives checkpoints, loss weights, and optimization settings.

\mypara{Control environment and policy.}
Our XLand environment adapts XLand-MiniGrid \citep{nikulin2023xlandminigrid}
with a \(9\times9\) room, local \(5\times5\times50\) observations,
six actions, rising-edge events, and a 75-step horizon.
The memoryless policy
has a linear observation projection and a shared three-layer torso of width 128.
\cref{fig:rl-environments} illustrates the environments.

\mypara{Comparison protocol.}
We compare against pretrained RAD embeddings \citep{yalcinkaya2024compositional}.
All policies use frozen encoders and true-DFA reward and termination;
residual embeddings are precomputed offline.
Conditioning follows
\cref{sec:decision-making}, with a task-blind control using \(g_t=0\).
Policies train with proximal policy optimization (PPO) \citep{schulman2017proximal}.
Each XLand policy trains for 16.384M environment steps, with five seeds.
Training pools contain either 25 tasks or 87/139/173 tasks for
RAD-compatible/multi-event/plain tasks, at fixed frames.
The RAD-compatible
pool contains tasks within the published RAD checkpoint's input limits.
We compare a zero
rejection reward, chosen by validation performance, with a rejection penalty
of $-1$.
Acceptance gives $+1$, and either verdict ends the episode in both
settings.
\cref{app:rl-protocol} records selection and evaluation protocols,
including test-informed development choices.

\subsection{Operator fidelity and composition}

The four-letter corpus has 4,014 presentations and 308 validation rows:
149 plain, 119 within-budget conjunctive, and 40 above \(C=16\).
Across five copied-formula seeds, all plain and within-budget conjunctive
rows agree with the input language on every word through length six.
The learned semantic model passes 139--143 plain and 114--115 conjunctive
rows across three seeds.
Copied-formula agreement verifies construction and coverage; retaining all
configurations within budget does not test learned ranking.
The copied construction preserves all within-bound compositions; both regimes
fail on most over-bound pairs.
\cref{tab:architecture,tab:restep} report architectural ablations and progress audits.
\cref{tab:composition} separately tests all-word equivalence on held-out
compositions, using component minimal DFAs for ground truth.

\begin{table}[t]
  \centering
  \caption{Composed operators accepting exactly the target language on all
    words, out of 30 held-out validation pairs per cell.
    Each column uses
    one checkpoint, with 16 operator slots and no composition-specific
    supervision.
    \(C_i\) counts reachable configurations of component
    automaton \(i\); the product/sum bounds can exceed the composed
    automaton's actual configuration count.}
  \label{tab:composition}
  \begin{tabular*}
    {\linewidth}{@{\extracolsep{\fill}}lccc@{}}
    \toprule
    & \multicolumn{3}{c}{Exactly equivalent compositions (out of 30)}\\
    \cmidrule(l){2-4}
    Formula assignment & Copied & Learned & Learned\\
    Transition construction & Syntactic & Semantic & Syntactic\\
    \midrule
    Conjunction, \(C_1C_2\leq16\)  & 30 & 25 & 20\\
    Disjunction, \(C_1+C_2\leq16\) & 30 & 29 & 20\\
    Conjunction, \(C_1C_2>16\)     & 6  & 6  & 5\\
    \bottomrule
  \end{tabular*}
\end{table}

\ifextendedpaper
\mypara{Representation geometry.}
\cref{fig:tsne-composition} visualizes validation task components and
compositions. These projections illustrate geometry but do not establish language
equivalence or semantic distance. The all-word composition audit above
provides the execution check.
\begin{figure}[t]
\centering
  \includestandalone[]{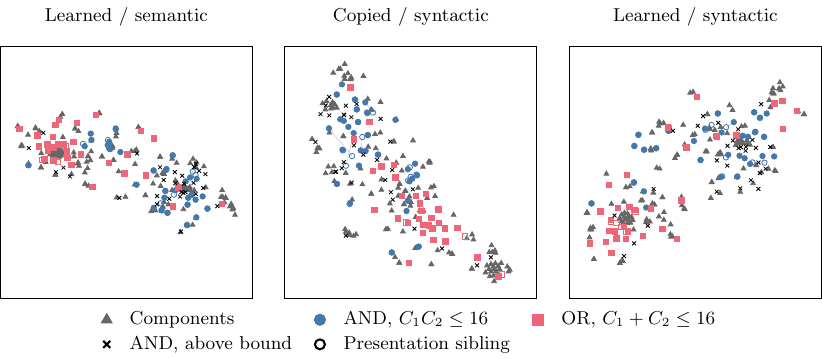}
\caption{t-SNE of validation components and compositions. Panels show learned
assignments/semantic transitions (\texttt{c3lsem-s0}), copied formulas/syntactic
transitions (\texttt{c3c-s4}), and learned assignments/syntactic transitions
(\texttt{c3lid-s0}). Gray triangles are components; circles/squares are
conjunctions/disjunctions, and crosses are conjunctions above the product
bound. \(C_i\) counts component configurations; operator capacity is 16.
Hollow markers replace a component by a language-equivalent presentation.
Axes are qualitative projection coordinates, not language distances.}
\label{fig:tsne-composition}
\end{figure}

\fi

\subsection{Task-conditioned control}

\cref{tab:rl-xland-preliminary} compares frozen semantic features and
RAD embeddings.
RAD-compatible and multi-event tasks require at least two
events for acceptance; the plain pool also includes single-event tasks.

\begin{table}[t]
  \centering\small
  \caption{Held-out XLand success after 16.384M steps with
    87/139/173 training tasks: means over five policy seeds, eight layouts, and 17/32/37 test tasks.
    RAD-compatible/Multi-event tasks require at least two events for acceptance,
    with minimal-DFA caps of 10/128 states.
    Plain uses non-product source tasks,
    including single-event objectives, with a 32-state DFA cap.
    Subcolumns give rejection reward; both settings terminate on rejection.
    Bold marks the highest mean per column; dashes denote untested arms.
    \cref{tab:xland-large-full} gives 95\%
    crossed seed/task/layout intervals.}
  \label{tab:rl-xland-preliminary}
  \begin{tabular*}{\linewidth}{@{\extracolsep{\fill}}lrrrrrr@{}}
\toprule
& \multicolumn{2}{c}{RAD-compatible} & \multicolumn{2}{c}{Multi-event} & \multicolumn{2}{c}{Plain}\\
Conditioning & \shortstack{Rejection\\reward $0$} & \shortstack{Rejection\\reward $-1$} & \shortstack{Rejection\\reward $0$} & \shortstack{Rejection\\reward $-1$} & \shortstack{Rejection\\reward $0$} & \shortstack{Rejection\\reward $-1$}\\
\midrule
Task-blind & 0.510 & 0.307 & 0.623 & 0.240 & 0.749 & 0.758\\
Ours/Initial & 0.560 & 0.429 & 0.634 & 0.505 & 0.775 & 0.771\\
Ours/Residual & 0.656 & 0.647 & 0.770 & \textbf{0.770} & 0.828 & 0.821\\
Ours/Carrier bits & 0.551 & 0.457 & 0.662 & 0.634 & 0.759 & 0.752\\
Ours/Carrier vectors & 0.653 & 0.640 & \textbf{0.773} & 0.755 & \textbf{0.833} & \textbf{0.833}\\
RAD/Initial & 0.516 & 0.434 & -- & -- & -- & --\\
RAD/Residual & \textbf{0.709} & \textbf{0.700} & -- & -- & -- & --\\
\bottomrule
\end{tabular*}

\end{table}

\ifextendedpaper
\mypara{Training trajectories.}
\cref{fig:xland-zero} shows XLand validation success
throughout policy training. Its pointwise intervals resample policy seeds;
the final test tables use crossed seed/task/layout resampling. Both views
retain all policy seeds. The additional XLand rejection-penalty and LetterWorld trajectories
appear in \cref{fig:xland-minus-one,fig:letterworld-final-curves}.
\begin{figure}[!t]
\centering%
  \includestandalone[]{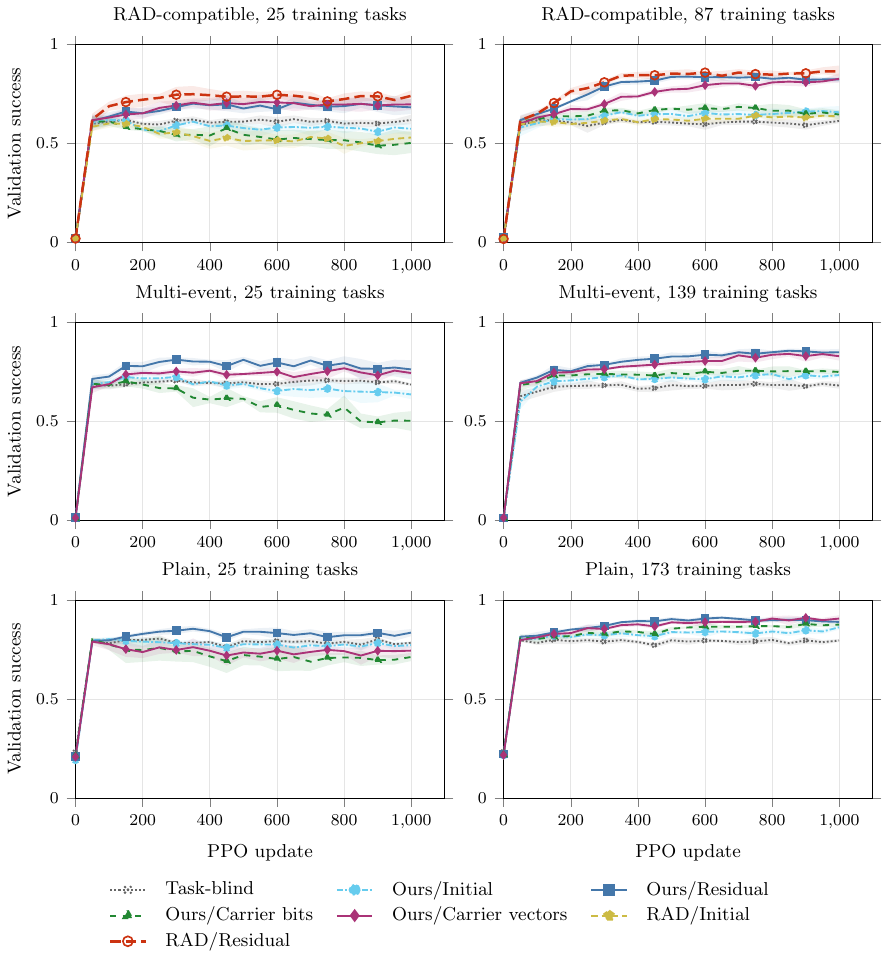}
\caption{XLand validation trajectories, rejection reward 0. Rows show task
pools; columns show 25 training tasks (left) and 87/139/173 training tasks
(right, for RAD-compatible/multi-event/plain). All five policy seeds are included. Bands are pointwise 95\% intervals
from 2,000 policy-seed bootstrap resamples. The selected architecture is linear/shared depth three.}
\label{fig:xland-zero}
\end{figure}

\fi

\mypara{Algebraic execution provides useful policy features.}
Pooled active-slot vectors outperform task-blind inputs and carrier bits in
all three large-pool conditions under both rewards, with difference intervals
above zero.
Multi-event success reaches 77.3\% under reward 0, an 11.0-point gain over
carrier bits.
This changes both feature content and input width (48 to 64), without
isolating pretraining from random or nonlearned features.

Carrier-vector success is within 1.5 percentage points of residual
re-encoding in every large-pool condition; both this comparison and the
matched RAD residual comparison remain inconclusive.
With only 25 training tasks, carrier vectors underperform task-blind inputs
on plain tasks under both rewards.
\cref{tab:xland-interface-differences,tab:xland-differences-small} report differences and confidence intervals.

\mypara{Compact execution extends beyond small DFAs.}
Fifteen of the 32 multi-event test tasks exceed the published RAD
checkpoint's ten-state DFA limit.
Using 16 operator slots, carrier vectors achieve 99.4\% and 96.9\% success
under rewards 0 and $-1$ on the four tasks with 33--128-state minimal DFAs.
This demonstrates compact control on those tasks; all size strata and their
uncertainty appear in \cref{tab:xland-size}.

\mypara{Transfer to LetterWorld.}
In LetterWorld \citep{vaezipoor2021ltl2action}, on RAD's task mix with the
corrected sampler, three policy seeds, and 20M
frames, our residual embeddings achieve 70.2\% success versus 30.8\% for
task-blind conditioning and 87.2\% for RAD.
The RAD advantage persists across two policy architectures.
These results demonstrate useful transfer with a remaining performance gap;
\cref{tab:letterworld-final,tab:letterworld-depth} give comparisons and post hoc depth diagnostics.

\section{Conclusion and Limitations}
\label{sec:conclusion}

Our algebraically informed AFA pipeline combines learned task and state
features with compact execution.
Configuration slots encode conjunctive obligations, and language coverage
ensures exact residual progression through Boolean matrix updates.
In XLand, pooling the active slots' learned vectors improves control over
carrier bits and task-blind inputs across large training pools under both
rewards, including tasks whose minimal DFAs exceed the slot budget.
LetterWorld shows useful transfer alongside a remaining performance gap.

\mypara{Limitations.}
\label{sec:limitations}
The experiments retain \(r=16\); they establish feasibility for compact
representations of large DFAs, without measuring scaling with operator budget
or controlling task difficulty across DFA sizes.
Pretraining and exact-inclusion compilation time and peak memory remain
unmeasured.
Policy learning remains sensitive to training-pool size, architecture, and
reward design.
Longer-path XLand pairs have low success despite full-state plans within the
horizon; the evaluated policies use local observations and no memory.
Exact task execution does not guarantee a successful control policy.
In LetterWorld, deep-residual vectors are weakly separated.
An exploratory fine-tune substantially reduces near-collisions without
a resolved control improvement, so this geometry alone does not explain the
observed gap.

\clearpage
\bibliography{main}

\ifincludeappendix
  \clearpage
  \appendix

\counterwithin{table}{section}
\renewcommand{\theHtable}{appendix.\thesection.\arabic{table}}

\section{Notation}
\label{app:notation}

\cref{tab:notation}
collects the symbols used in the construction.
All operator products are taken
in the stated semiring; hidden-vector maps and training losses use real
arithmetic.
Vectors \(\vect{\alpha}\) and \(\vect{\beta}\) are columns;
\(\vect{c}_u\) is a row.
We omit a batch axis from mathematical expressions.

\mypara{Matrix operators.}
We use Boolean matrix arithmetic, with \(\Be=\{0,1\}\),
\((AB)_{ij}=\bigvee_k(A_{ik}\land B_{kj})\), and Boolean identity \(I\).

For the configuration NFA of \(\Ac\), we enumerate the reachable
configurations as \(\tau_1,\ldots,\tau_C\) and define its transition
matrices by
\begin{equation}
  M_{\Ac,a}\in\Be^{C\times C},\qquad
  M_{\Ac,a}[i,j]=\mathbf{1}[\tau_i\xrightarrow{a}\tau_j].
\end{equation}
Let \(\alpha_{\Ac,i}=\mathbf{1}[\tau_i\in[\iota]]\) indicate initial
configurations and \(\beta_{\Ac,i}=\mathbf{1}[\tau_i\subseteq F]\) indicate
accepting configurations, giving the matrix representation of \(\Ac\):
\begin{equation}
  \Tuple{\vect{\alpha}_\Ac,(M_{\Ac,a})_{a\in\Sigma},
    \vect{\beta}_\Ac},\qquad
  \vect{\alpha}_\Ac,\vect{\beta}_\Ac\in\Be^C.
  \label{eq:operator}
\end{equation}
Define the word-transition map \(M_\Ac:\Sigma^*\to\Be^{C\times C}\)
by \(M_\Ac(a_1\cdots a_k)=M_{\Ac,a_1}\cdots M_{\Ac,a_k}\), with
\(M_\Ac(\varepsilon)=I_C\).

We take both indicator vectors as columns and represent the active set
of NFA states through a Boolean row vector, which we call a
\emph{carrier}.
We then execute the NFA and evaluate word acceptance by
\begin{equation}
  \begin{aligned}
    \vect{c}_{\Ac,\varepsilon} & =\vect{\alpha}_\Ac^{\mathsf T}, &
    \vect{c}_{\Ac,ua}          & =\vect{c}_{\Ac,u}M_{\Ac,a},       \\
    f_\Ac(w)                   & =\vect{\alpha}_\Ac^{\mathsf T}
    M_\Ac(w)\vect{\beta}_\Ac.
  \end{aligned}
  \label{eq:carrier-appendix}
\end{equation}
We accept \(w\) exactly when \(f_\Ac(w)=1\), so
\(L(\Ac)=\Set{w\in\Sigma^*\given f_\Ac(w)=1}\).
Throughout, we omit the subscript \(\Ac\) when the underlying automaton
is clear from context.

For an \(r\)-state NFA, the same construction uses vectors in \(\Be^r\)
and matrices in \(\Be^{r\times r}\).

\mypara{Semiring arithmetic.}
NFA execution admits a matrix formulation over a semiring.
A semiring \(\Ke=\Tuple{K,\oplus,\otimes,\eplus,\etimes}\) has associative
addition and multiplication, commutative addition, their respective
identities, distributivity, and an annihilating additive identity
\citep{golan1999semirings}.
For matrices of compatible dimensions, these operations extend as
addition entrywise and multiplication through sums of products:
\begin{equation}
  (A\oplus B)_{ij}=A_{ij}\oplus B_{ij},\qquad
  (AB)_{ij}=\bigoplus_k A_{ik}\otimes B_{kj}.
\end{equation}
For square matrices, these operations give the matrix semiring
\(K^{r\times r}\).
We use the matrix with every entry equal to \(\eplus\)
as its additive identity and the matrix \(I_r\), with diagonal entries
\(\etimes\) and off-diagonal entries \(\eplus\), as its multiplicative
identity.
We specialize to the Boolean semiring
\(\Tuple{\Be=\{0,1\},\lor,\land,0,1}\), so that matrix multiplication
records the existence of transition paths.

\begin{table}[htbp]
  \centering
  \caption{Automaton, encoder, and execution notation for one task; batch
    axes are omitted. AFA/DFA mean alternating/deterministic finite automaton,
    and \(\Be=\{0,1\}\). Initial and accepting vectors are columns;
    carriers are rows.}
  \label{tab:notation}
  \small
  \begin{tabular}{lp{.65\linewidth}}
    \toprule
    Symbol & Meaning and domain\\
    \midrule
    \(\Ac,\Sigma,Q,F\) & Input AFA, event alphabet, source states, accepting states.\\
    \(n,m\) & Source-state count \(\abs Q\) and alphabet size \(\abs\Sigma\).\\
    \(\PBool[Q],\iota,\Delta\) & Positive Boolean formulas; initial formula and transition map \(Q\times\Sigma\to\PBool[Q]\).\\
    \([p]\) & Subsets of \(Q\) appearing as terms in the irredundant disjunctive normal form of \(p\).\\
    \(u,v,w,\varepsilon\) & Event words and empty word; \(\abs u\) is length.\\
    \(L_\Ac(p),u^{-1}L\) & Language with initial formula \(p\); residual \(\{v:uv\in L\}\).\\
    \(\Ac_u\) & Input AFA with initial formula \(\Delta_u(\iota)\).\\
    \(\tau,\Reachable(\Ac),C\) & Configuration \(\tau\subseteq Q\), reachable configuration family, and its cardinality.\\
    \(D_\Ac,\abs{D_\Ac}\) & Minimal DFA for \(L(\Ac)\) and its number of states.\\
    \(V,E,v_\iota\) & Hypergraph vertices \(Q\sqcup\{v_\iota\}\), hyperedges, and fresh initial vertex.\\
    \(T_e,S_{q,a}\) & Targets of hyperedge \(e\); edges leaving \(q\) under event \(a\).\\
    \(x_v,x_e\) & Vertex features in \(\Be^2\) and edge features in \(\Be^{m+1}\).\\
    \(L,d,d_\pi\) & Encoder depth, vertex-vector width, and pinned-vector width.\\
    \(h_v^{(\ell)},z_\Ac\) & Layer-\(\ell\) vertex vector and initial-vertex task readout, in \(\Re^d\).\\
    \(\vect\pi_q\) & Pinned source-state vector in \(\Re^{d_\pi}\).\\
    \(E_\theta,D_\eta,\mathcal H_\Ac\) & Encoder, operator decoder, and structured output: task/state vectors, formula factors, and acceptance data.\\
    \(r,N\) & Operator-slot budget and candidate cap (16 and 64 in the main experiments).\\
    \(J,\tau_t,\ell_t\) & Live-slot indices, selected configuration, and language \(\ell_t=L_\Ac(\tau_t)\).\\
    \(M_{\Ac,a},M_\Ac(w)\) & Event matrix and ordered word-transition product in \(\Be^{r\times r}\).\\
    \(\vect\alpha,\vect\beta\) & Initial and accepting indicators in \(\Be^{r\times1}\).\\
    \(\vect c_u\) & Active-slot carrier after \(u\), in \(\Be^{1\times r}\).\\
    \(f_\Ac(w)\) & Boolean acceptance score \(\vect c_w\vect\beta\).\\
    \(\vect v_j,\overline{\vect v}(\vect c)\) & Mean pinned vector of configuration \(j\), and mean over active slots; both in \(\Re^{d_\pi}\).\\
    \bottomrule
  \end{tabular}
\end{table}

\begin{table}[htbp]
  \centering
  \caption{Assignment, training, and policy notation. Plain \(g\) indexes
    a formula; \(g_t\) is a policy input. Plain \(c_{ek}\) is a copied
    conjunct symbol; bold \(\vect c_u\) is the Boolean carrier.
    Indices are local to their definitions.}
  \label{tab:notation-learning}
  \small
  \begin{tabular}{lp{.65\linewidth}}
    \toprule
    Symbol & Meaning and domain\\
    \midrule
    \(\theta,\eta,\psi\) & Shared encoder, decoder, and policy parameters.\\
    \(\mathcal F_\eta,\widehat\iota,\widehat\Delta\) & Learned formula-assignment map and its initial/transition formulas.\\
    \(g,E_g\) & Formula index \(g\in(Q\times\Sigma)\sqcup\{\iota\}\) and eligible input hyperedges.\\
    \(R,K\) & Numbers of disjunct and conjunct positions in each decoded formula.\\
    \(w_{gje}\) & Nonnegative weight assigning edge \(e\) to disjunct \(j\); sums to one over eligible edges plus the false choice.\\
    \(c_{ek}\) & Copied target symbol in conjunct position \(k\) of edge \(e\), padded with \(\top\).\\
    \(P_{gjk}(x)\) & Mixed factor weight for symbol \(x\in Q\sqcup\{\top,\bot\}\).\\
    \(\widehat\tau_{gj},\widehat\psi_g\) & Thresholded conjunctive term and decoded formula.\\
    \(\mathcal C_N,s_\eta,\operatorname{Top}_k\) & Enumerated candidate family, learned candidate score, and selection of \(k=\min(r,\abs{\mathcal C_N})\) live slots.\\
    \(\widetilde\alpha,\widetilde M,\widetilde\beta\) & Training-time soft arrays with entries in \([0,1]\).\\
    \(\epsilon_\ell,T\) & Encoder pooling temperature and operator smoothing temperature.\\
    \(s_\Ac(w),y_\Ac(w)\) & Soft word score in \([0,1]\) and exact membership label in \(\{0,1\}\).\\
    \(\Lc,\lambda_k\) & Combined supervision objective and nonnegative weight for loss term \(k\).\\
    \(\rho,m_{\mathrm{id}},j\) & Identity-loss discount, base margin, and shortest distinguishing-word length.\\
    \(o_t,u_t,e_t\) & Physical observation, control action, and emitted event word \(e_t\in\Sigma^*\).\\
    \(w_t,g_t,\pi_\psi\) & Accumulated event prefix, conditioning vector, and shared policy \(\pi_\psi(u_t\mid o_t,g_t)\).\\
    \bottomrule
  \end{tabular}
\end{table}

The same letter \(u\) is used for a word when written without a physical time
index and for a control action in \(u_t\); its domain is explicit at use.
The \(D/B\) report columns are written as
\(N_6^{\mathrm{dec}}/N_6^{\mathrm{true}}\) in the paper, to distinguish counts
from the minimal DFA \(D_\Ac\).
Layer temperatures are not configurations.
The parameter \(\rho\) in the identity loss is distinct from the policy
discount \(\gamma\).

\begin{table}[htbp]
  \centering
  \caption{Names used in the policy-conditioning tables and curve legends.
    Ours/m denotes Morphata; RAD/r denotes the comparison encoder.
    Encoders are frozen, and all listed arms use true-DFA reward and
    termination. Carrier interfaces additionally retain the initial embedding.}
  \label{tab:rl-vocabulary}
  \small
  \begin{tabular}{llp{.43\linewidth}}
    \toprule
    Display name & Run/legend code & Conditioning input\\
    \midrule
    Task-blind & \texttt{none} & Zero vector.\\
    Ours/Initial & \texttt{m-init} & Fixed initial Morphata embedding.\\
    RAD/Initial & \texttt{r-init} & Fixed initial RAD embedding.\\
    Ours/Residual & \texttt{m-res} & Morphata embedding of the current residual language.\\
    RAD/Residual & \texttt{r-res} & RAD embedding of the current residual language.\\
    Ours/Carrier bits & \texttt{m-op} & Initial embedding concatenated with Boolean active-slot indicators.\\
    Ours/Carrier vectors & \texttt{m-opv} & Initial embedding concatenated with the mean active-slot vector.\\
    \bottomrule
  \end{tabular}
\end{table}

\section{Worked Example: Conjunctive Obligations and Residual Execution}
\label{app:running-example}

Consider the task of observing both $a$ and $b$, in either order. We construct
its configuration and semantic operators, follow their carriers along $ba$,
and compute the corresponding policy features. All four reachable
configurations are retained; the example illustrates execution and pooling
without assuming learned selection or particular encoder outputs.

\mypara{Task automaton.}
Let $\Sigma=\{a,b\}$, $Q=\{q_a,q_b\}$, $\iota=q_a\wedge q_b$, and
$F=\emptyset$. The transition formulas are
\[
  \Delta(q_a,a)=\top,\qquad \Delta(q_a,b)=q_a,\qquad
  \Delta(q_b,a)=q_b,\qquad \Delta(q_b,b)=\top.
\]
Writing $L_x=\Sigma^*x\Sigma^*$ for $x\in\Sigma$, we have
$L_\Ac(q_a)=L_a$, $L_\Ac(q_b)=L_b$, and $L(\Ac)=L_a\cap L_b$.
The empty word is rejected. After $b$, the initial formula progresses to
$q_a$; after $ba$, it is $\top$ and the task is accepted.

The hypergraph has vertices $Q\sqcup\{v_\iota\}$ and edges
\[
  E=\{(q_a,\emptyset,a),(q_a,\{q_a\},b),
       (q_b,\{q_b\},a),(q_b,\emptyset,b),
       (v_\iota,\{q_a,q_b\},\mathrm{init})\}.
\]
The initial edge groups the two obligations conjunctively. Edges with empty
target sets encode $\top$. The encoder supplies $z_\Ac$ and source-state
vectors $\vect\pi_{q_a},\vect\pi_{q_b}$ from this structure.

\mypara{Configuration closure.}
Exploration from $[\iota]=\{\{q_a,q_b\}\}$ yields
\[
 \tau_1=\{q_a,q_b\},\quad \tau_2=\{q_b\},\quad
 \tau_3=\{q_a\},\quad \tau_4=\emptyset.
\]
Their languages are $\ell_1=L_a\cap L_b$, $\ell_2=L_b$, $\ell_3=L_a$,
and $\ell_4=\Sigma^*$. Thus $C=4$, and $r\geq4$ retains the complete
closure. Additional slots are zero padding. The empty configuration is an
accepting slot: its conjunction is $\top$, and $\emptyset\subseteq F$.
An empty carrier, in contrast, represents the empty language.

\begin{figure}[tb]
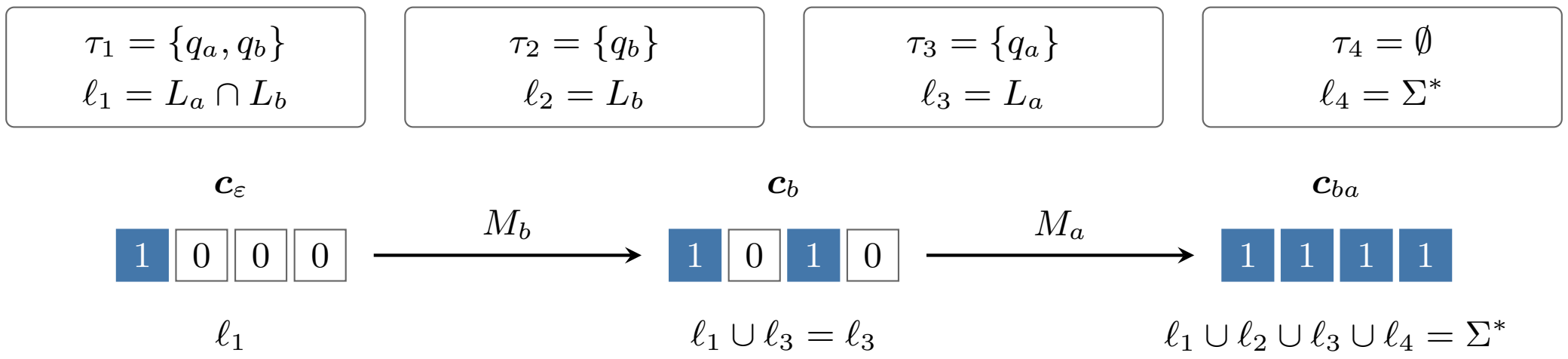

  \centering
  \includestandalone[]{figures/standalone/execution-example}
  \caption{Semantic execution for the task of observing both $a$ and $b$.
    Each top box gives a configuration $\tau_t$ and its language $\ell_t$,
    with $L_x=\{w\in\{a,b\}^*:w\text{ contains }x\}$.
    The bottom row follows the carrier along $ba$, using Boolean matrix
    products. Entries are ordered by slots $1$--$4$, with active entries
    filled. The language below each carrier is the union of its active
    configuration languages. Multiple active slots can represent the same
    residual as a single slot.}
  \label{fig:example-semantic-trace}
\end{figure}

\mypara{Configuration-NFA operator.}
Let $N_a,N_b$ denote the matrices obtained by taking the irredundant DNF
terms of each successor formula. Using the slot order above,
\[
 N_a=\begin{pmatrix}0&1&0&0\\0&1&0&0\\0&0&0&1\\0&0&0&1\end{pmatrix},
 \qquad
 N_b=\begin{pmatrix}0&0&1&0\\0&0&0&1\\0&0&1&0\\0&0&0&1\end{pmatrix},
 \qquad
 \vect\alpha=\begin{pmatrix}1\\0\\0\\0\end{pmatrix},\quad
 \vect\beta=\begin{pmatrix}0\\0\\0\\1\end{pmatrix}.
\]
Rows index source slots and columns index destination slots. For example,
$\tau_1$ progresses to $\tau_3$ on $b$, then to $\tau_4$ on $a$.
The carrier therefore follows
$(1,0,0,0)\xrightarrow{b}(0,0,1,0)\xrightarrow{a}(0,0,0,1)$.

\mypara{Semantic and syntactic operators.}
The semantic rule $M_a[t,t']=\mathbf1[\ell_{t'}\subseteq a^{-1}\ell_t]$
includes every retained language contained in the successor residual.
Here it gives
\[
 M_a=\begin{pmatrix}1&1&0&0\\1&1&0&0\\1&1&1&1\\1&1&1&1\end{pmatrix},
 \qquad
 M_b=\begin{pmatrix}1&0&1&0\\1&1&1&1\\1&0&1&0\\1&1&1&1\end{pmatrix},
\]
with the same $\vect\alpha$ and $\vect\beta$. Copied-formula syntactic
construction gives these same matrices for this task: for example, both
$\tau_1$ and $\tau_2$ satisfy $\Delta(q_a,a)\wedge\Delta(q_b,a)=q_b$.
Equality of the two constructions here is specific to this example.

The semantic carrier follows
$(1,0,0,0)\xrightarrow{b}(1,0,1,0)\xrightarrow{a}(1,1,1,1)$,
as shown in \cref{fig:example-semantic-trace}.
After $b$, its language is $\ell_1\cup\ell_3=\ell_3=b^{-1}L(\Ac)$;
after $ba$, it is $\Sigma^*$. Initial coverage holds because
$L(\Ac)=\ell_1$. Successor coverage holds because every one-event residual
of each $\ell_t$ is itself a retained slot language. Both carriers therefore
track the residual exactly, and both accept precisely when slot $4$ is active.

\mypara{Policy features.}
The slot vectors are
\[
 \vect v_1=\tfrac12(\vect\pi_{q_a}+\vect\pi_{q_b}),\quad
 \vect v_2=\vect\pi_{q_b},\quad \vect v_3=\vect\pi_{q_a},\quad
 \vect v_4=\vect0.
\]
Averaging the active slot vectors gives the features in
\cref{tab:example-pooling}. The initial vector $z_\Ac$ remains fixed in
both $[z_\Ac,\vect c_u]$ and $[z_\Ac,\overline{\vect v}(\vect c_u)]$.
Residual conditioning after $b$ uses $z_{\Ac_b}$, where $\Ac_b$ has initial
formula $q_a$.

\begin{table}[tb]
 \centering\small
 \caption{Pooled control features along $ba$ for the task of observing both
 $a$ and $b$. Each entry is the mean of active slot vectors; source-state
 vectors $\vect\pi_{q_a},\vect\pi_{q_b}$ are arbitrary encoder outputs.
 The configuration-NFA matrices are $N_a,N_b$; the semantic matrices are
 $M_a,M_b$, which also equal the copied-formula syntactic matrices here.
 Both operators represent the residual in the second column exactly.}
 \label{tab:example-pooling}
 \begin{tabular}{@{}cccc@{}}
 \toprule
 Prefix $u$ & Residual language & Configuration-NFA pooling & Semantic pooling\\
 \midrule
 $\varepsilon$ & $L_a\cap L_b$ &
 $\tfrac12(\vect\pi_{q_a}+\vect\pi_{q_b})$ &
 $\tfrac12(\vect\pi_{q_a}+\vect\pi_{q_b})$\\[3pt]
 $b$ & $L_a$ & $\vect\pi_{q_a}$ &
 $\tfrac14(3\vect\pi_{q_a}+\vect\pi_{q_b})$\\[3pt]
 $ba$ & $\Sigma^*$ & $\vect0$ &
 $\tfrac38(\vect\pi_{q_a}+\vect\pi_{q_b})$\\
 \bottomrule
 \end{tabular}
\end{table}

Exact residual execution does not make the pooled features canonical.
After $ba$, configuration-NFA pooling is zero, while semantic pooling is
generally nonzero. These carriers represent the same language through
different active configurations. Likewise, slot indices have task-specific
meanings: relabeling $a,b$ as $c,d$ preserves the carrier trace while changing
the obligations it denotes. Pooling exposes the associated learned state
vectors, whose quality for control remains an empirical question.

\section{Extended Proofs}
\label{app:proofs}

\subsection{Proof of exact residual execution}
\label{app:residual-proof}

We prove the residual-execution theorem from the main paper. Let \(J\) be
the live slots, with configuration languages \(\ell_t\), and use the
semantic indicators
\begin{equation}
  \alpha_t=\mathbf{1}[\ell_t\subseteq L(\Ac)],\qquad
  M_a[t,t']=\mathbf{1}[\ell_{t'}\subseteq a^{-1}\ell_t],\qquad
  \beta_t=\mathbf{1}[\varepsilon\in\ell_t]
\end{equation}
for live indices, with all padded entries zero. We assume the coverage
identities
\begin{align}
  L(\Ac)&=\bigcup_{\substack{t\in J\\\ell_t\subseteq L(\Ac)}}\ell_t,
  \nonumber\\
  a^{-1}\ell_t&=
  \bigcup_{\substack{t'\in J\\\ell_{t'}\subseteq a^{-1}\ell_t}}\ell_{t'},
  \qquad t\in J,\ a\in\Sigma.
\end{align}

\begin{proof}
We show by induction on \(u\) that
\begin{equation}
  u^{-1}L(\Ac)=\bigcup_{\substack{t\in J\\c_u[t]=1}}\ell_t.
  \label{eq:app-carrier-invariant}
\end{equation}
For \(u=\varepsilon\), the initial carrier is
\(\vect{c}_\varepsilon=\vect{\alpha}^{\mathsf T}\). Its active slots
are exactly those whose languages are contained in \(L(\Ac)\), so initial
coverage proves the identity.

Suppose the identity holds for \(u\). Boolean matrix multiplication gives
\(c_{ua}[t']=1\) exactly when some active slot \(t\) has \(M_a[t,t']=1\).
Consequently,
\begin{align}
  \bigcup_{\substack{t'\in J\\c_{ua}[t']=1}}\ell_{t'}
  &=\bigcup_{\substack{t\in J\\c_u[t]=1}}
    \ \bigcup_{\substack{t'\in J\\\ell_{t'}\subseteq a^{-1}\ell_t}}
    \ell_{t'}\nonumber\\
  &=\bigcup_{\substack{t\in J\\c_u[t]=1}}a^{-1}\ell_t\nonumber\\
  &=a^{-1}\!\left(\bigcup_{\substack{t\in J\\c_u[t]=1}}\ell_t\right)
    =a^{-1}(u^{-1}L(\Ac))=(ua)^{-1}L(\Ac).
\end{align}
The second equality uses successor coverage, and the last uses the induction
hypothesis. This proves the carrier interpretation for every finite word.
\end{proof}

For the acceptance-correctness corollary, we use the definition of
\(\vect{\beta}\):
\begin{equation}
  \begin{aligned}
    \vect{c}_u\vect{\beta}=1
    &\ \Longleftrightarrow\
    \varepsilon\in\bigcup_{\substack{t\in J\\c_u[t]=1}}\ell_t\\
    &\ \Longleftrightarrow\
    \varepsilon\in u^{-1}L(\Ac)
    \ \Longleftrightarrow\ u\in L(\Ac).
  \end{aligned}
\end{equation}
The argument includes an empty live-slot family when \(L(\Ac)=\emptyset\).

\subsection{Configuration languages and capacity}

Fix an AFA and write \(\Delta_a(\tau)\) for
\(\bigwedge_{q\in\tau}\Delta(q,a)\). For every configuration and event,
\begin{equation}
 a^{-1}L(\tau)=L(\Delta_a(\tau))
   =\bigcup_{\tau'\in[\Delta_a(\tau)]}L(\tau').
 \label{eq:app-configuration-recursion}
\end{equation}
The first equality follows from substitution in the acceptance definition.
The second expands a positive formula into its disjunctive normal form.
At the empty word, acceptance is \(\tau\subseteq F\). Thus induction on word
length proves that the NFA with configurations as states, the initial terms
of \(\iota\), and these transitions accepts exactly \(L(\Ac)\).
This argument includes the empty configuration, whose language is
\(\Sigma^*\), and an empty initial term family, whose language is empty.

To check the syntactic operator, note that
\(\tau'\models\Delta_a(\tau)\) implies
\(L(\tau')\subseteq a^{-1}L(\tau)\). Indeed, a positive formula is true
under a state set precisely when that set contains at least one of its prime
implicants; intersection over a larger state set can only reduce a language.
The same argument gives
\(\tau\models\iota\Rightarrow L(\tau)\subseteq L(\Ac)\).
Every actual prime-implicant successor and every actual initial term is
included in these tests. If all reachable configurations occupy slots, the
operator contains the configuration NFA's accepting runs and adds only
language-sound choices. This proves exactness of the syntactic construction.
Semantic inclusion contains these sound edges and initials, so the same
soundness and completeness argument applies to it. The two matrices need
not be entrywise identical.

If \(C\leq r\), place the \(C\) configurations in slots and pad the rest
inertly. The initial configurations cover \(L(\Ac)\), and every successor
term of a reachable configuration is reachable, so the configuration recursion
proves both coverage conditions. This proves the sufficient-capacity corollary. It is independent of
how the configurations are obtained and of whether a learned selector finds
them. When closure enumeration produces at most \(r\) candidates and top-\(r\)
retains every live candidate, copied-formula coverage is structural.
Conversely, \(C>r\) is not a lower bound on every NFA for the language.
A different generating family may fit the operator.

\subsection{Semantic closure and language sharing}

Let \(\ell_1,\ldots,\ell_r\) be the selected configuration languages. Suppose
that initial coverage and letter-quotient closure hold as in the main paper.
Starting in slot \(t\), let \(G_t\) be the language of the semantic operator.
We show by induction on \(\abs w\) that
\(w\in G_t\Longleftrightarrow w\in\ell_t\).
For \(w=\varepsilon\), this is the definition of \(\beta_t\).
For \(w=av\), the operator accepts exactly when some \(t'\) satisfies
\(\ell_{t'}\subseteq a^{-1}\ell_t\) and \(v\in G_{t'}\).
By the induction hypothesis and closure, this is equivalent to
\(v\in a^{-1}\ell_t\), hence to \(av\in\ell_t\).
Initial coverage then identifies the language of the full operator with
\(L(\Ac)\).

Take the family of all reachable configuration languages. Equation
\eqref{eq:app-configuration-recursion} gives quotient closure, and the
initial configurations cover \(L(\Ac)\). Removing duplicate languages changes
neither union. Consequently one representative for each distinct language
suffices. This is an existence result about the semantic construction; it
does not establish exact equality of learned vectors, nor does it show that
top-\(r\) selection will choose representatives.

\subsection{Residuals and carrier distances}

For an operator accepting \(L\), associativity gives
\begin{equation}
 \vect{\alpha}^{\mathsf T}M(uv)\vect{\beta}
 =\vect{\alpha}^{\mathsf T}M(u)M(v)\vect{\beta}
 =\vect{c}_uM(v)\vect{\beta}.
\end{equation}
Hence the stepped triple accepts \(u^{-1}L\). The same calculation shows
that concatenation is represented by matrix products for every operator,
independently of its training or acceptance accuracy.
If two prefixes give the same carrier, every continuation has the same
operator acceptance. Distinct carriers may still have identical continuation
languages. Comparing carriers from different operators requires their fixed
matrices and final weights.

For an exact task operator, the set of accepting continuations at every
reachable prefix equals the true residual language. Taking the minimum word
length in this set gives the same completion distance in the carrier graph
and minimal DFA. If the set is empty, neither has a finite distance.
A horizon-limited search only certifies the absence of an accepting word up
to that horizon. Event restrictions and no-op symbols must agree between the
two run graphs for this distance comparison.

\subsection{Conjunction and disjunction}

Rename component states to be disjoint. Since transitions of each component
refer only to its own states, substitution commutes with conjunction and
disjunction of the initial formulas. Evaluating the resulting formula at
\(F_1\sqcup F_2\) proves intersection and union of the component languages.
For disjunction, a reachable configuration comes from one component, giving
at most \(C_1+C_2\) possibilities. For conjunction, it is a union
\(\tau_1\cup\tau_2\) of component configurations, giving at most
\(C_1C_2\) possibilities. Formula simplification can reduce these counts.
In particular, a product bound exceeding \(r\) does not prove that the
composed task requires more than \(r\) slots.

\subsection{Permutation conventions}

The shared AND-OR encoder maps isomorphic typed incidence graphs to
correspondingly permuted vertex vectors; the initial-vertex task vector is
unchanged. This follows layer by layer from shared maps and order-independent
pools. Pinned vectors retain the ordering of source states.

For a fixed compiled operator and slot-permutation matrix \(P\), define
\begin{equation}
 \vect{\alpha}'=P^{\mathsf T}\vect{\alpha},\qquad
 M'_a=P^{\mathsf T}M_aP,\qquad
 \vect{\beta}'=P^{\mathsf T}\vect{\beta}.
\end{equation}
Its carrier is \(\vect{c}'_u=\vect{c}_uP\), and word acceptance is unchanged.
This coordinate identity does not establish that independently compiled
presentations select the same languages, or that formula-head truncation and
ordering yield a globally equivariant decoder. Those are separate
implementation and evaluation questions.

\subsection{Configuration reachability and size}

Each NFA state thus records simultaneous obligations, while the active
set of NFA states records alternatives.
We let \({\to}=\bigcup_{a\in\Sigma}{\xrightarrow{a}}\) collect the
configuration transitions and use its reflexive transitive closure
\(\to^*\) to define
\begin{equation}
  \Reachable(\Ac)=
  \Set{\tau\subseteq Q\given
    \exists\tau_0\in[\iota],\ \tau_0\to^*\tau},
  \qquad C=\abs{\Reachable(\Ac)}.
\end{equation}
Thus, we count configurations reachable from the initial configurations
across all finite words, including the empty word.
Since each configuration is a subset of \(Q\), this NFA has at most
\(C\leq 2^{\abs{Q}}\) states, allowing an exponential increase over the
input AFA.
The configuration NFA accepts \(L(\Ac)\); determinizing it
introduces a further exponential bound of \(2^C\) states for the minimal
DFA, which we denote by \(D_\Ac\).
Together, these constructions give a
doubly exponential upper bound of \(2^{2^{\abs{Q}}}\) relative to the AFA
state count.

\section{Architecture and Compilation Details}
\label{app:architecture}

\subsection{Interfaces and dimensions}

For one automaton, let \(n=\abs Q\), \(m=\abs\Sigma\), and \(b=\abs E\).
The incidence representation has \(n+1\) vertex nodes, including the initial
vertex. Vertex features have width two: acceptance and root indicators.
Finite-alphabet hyperedge features have \(m\) guard-membership columns and
one initial-edge column. Source incidences associate each hyperedge with its
source, and target incidences associate it with every required state.
Parallel hyperedges and self-targeting edges retain their incidences.

After each of the eight adopted layers, vertex and hyperedge states have
width 32. For each layer, the state update reads \(1+32m\) features before
its neural map: one accepting bit and one pooled vector per letter.
After updating the source states, the initial vertex pools their new values
through its initial hyperedges and applies its separate residual map and
layer normalization. The initial task vector has width 32.

The pinned-state array has 16 rows of width 32; rows beyond the input's states
are zero and treated as invalid. The structured encoder output concatenates
the task vector, pinned vectors, transition factors, initial factors, and
accepting bits. A transition-factor tensor has shape
\(m\times16\times8\times6\times18\): symbols, source slots, disjunct
positions, conjunct positions, and choices. The choices are true, false, and
the 16 state slots. Initial factors have shape \(8\times6\times18\).
These factor shapes describe the adopted experimental head; they are not
the shape of the policy's input.

The graph-to-factor head retains at most 64 input hyperedges at four letters.
The conjunction cap is six target positions per factor term and the disjunction
cap is eight positions. These limits are separate from the 16 output
configuration slots and the 64 enumerated closure candidates. Any omitted
input edge or truncated target set can affect the candidate construction.

The active hyperedge caps are 64/128/256 at four/ten/sixteen letters,
respectively. These are the experimental head's \texttt{EDGE\_CAP} values;
the library specification's separate default field does not determine this
truncation. All configurations use 16 pinned source-state vectors and at
most 64 closure candidates. Pretraining inputs have at most eight source
states; edge and formula caps remain distinct restrictions.

\subsection{Formula heads and closure selection}
\label{app:formula-selection}

For copied formulas, factors are filled from the input's hyperedges and
padded with the appropriate constants. In the partially learned regime,
conjunct positions copy the members of an input hyperedge. A learned head
assigns hyperedges to transition disjunct positions, with an exclusive
assignment option. The initial formula has its own assignment head.
Transition weights are masked to edges with the matching source and letter;
initial weights are masked to initial edges. Both heads include a learned
false-choice logit and normalize over eligible edges plus that choice.
When the exclusivity option is enabled, transition weights undergo three
passes of edge-column rescaling and row normalization: each edge column is
divided by the maximum of one and its sum over disjunct positions, then each
row is normalized. The false column is exempt from column rescaling.
The initial head uses its own softmax weights without this adjustment.
The resulting weights mix copied conjunct indicators before thresholding;
they do not implement a discrete one-edge choice per disjunct.

Acceptance comes from the input's accepting bits. The precise head behavior
for the initial formula must be distinguished from transition assignments
when tracing failures.

Thresholded factors determine conjunction membership and whether a disjunct
is live. Breadth-first closure starts from the decoded initial terms. Each
candidate's successor terms are inserted until no new candidate appears or
the cap \(N=64\) is reached. Candidates are therefore reachable under the
thresholded formulas. This is not a differentiable enumeration.
For a candidate \(\tau_j\) discovered at depth \(b_j\), the adopted selection
head reads
\begin{equation}
 s_j=\operatorname{MLP}_\eta\!\left(
  \left[\sum_{q\in\tau_j}\vect{\pi}_q,\ \abs{\tau_j},\ b_j/8\right]\right).
 \label{eq:app-selection}
\end{equation}
The selected configurations are the top \(r\) candidates by \(s_j\).
A softmax surrogate supplies gradients to the scores while the forward pass
uses hard choices. Invalid candidates are masked. When all live candidates
fit in the operator, every live candidate is retained regardless of ranking.

\subsection{Two kinds of smooth pooling}

The weighted max-min encoder variant gates an AND member \(x\) as
\(\max(1-g,x)\) and an OR member as \(\min(w,x)\), with learned
coordinatewise \(g,w\in[0,1]^d\). This is an architectural alternative
to the AND-OR encoder used in the adopted configuration. Encoder pooling
temperatures are learned subject to the configured floor.

The encoder uses the log-sum-exp extremes in the main paper. The operator's
smooth reductions instead use the softmax-weighted average
\begin{equation}
 \operatorname{ext}_{T,\sigma}(x_1,\ldots,x_k)
   =\sum_i\frac{\exp(\sigma x_i/T)}{\sum_j\exp(\sigma x_j/T)}x_i,
 \qquad \sigma\in\{-1,1\}.
 \label{eq:app-operator-extreme}
\end{equation}
It lies between the smallest and largest inputs and converges to the
corresponding extreme as \(T\) tends to zero. The adopted operator smoothing
temperature is \(0.03\). These reductions and the straight-through gradients
for word-level max/min are distinct operations in the computation graph.

\mypara{Stable masked encoder pooling.}
The first implementation overflowed while exponentiating rows outside the
selected pool. Its first six AND-OR runs became non-finite between steps
7,000 and 19,000. Masking those exponentials afterward hid the forward infinity
but allowed a zero-times-infinity product in the backward pass. For a pool
\(R\), the corrected reduction uses a stopped-gradient maximum
\(p=\max_{i\in R}(\sigma x_i/\epsilon)\), masks the shifted scores before
exponentiation, and sums only selected rows. Every selected exponent is then
nonpositive. Empty pools have an explicit learned value. This removes the
diagnosed overflow mechanism without requiring a positive temperature floor
for that purpose. The rerun completed 75k steps in all nine runs. Non-finite
losses or gradient norms cause the training loop to stop.

\subsection{Semantic inclusion and compilation}

Write \(B_q(w)=\mathbf{1}[w\in L(q)]\). Given a possibly soft configuration
indicator \(T_{tq}\), its bounded word table uses the smooth counterpart of
\begin{equation}
 \widetilde\ell_t(w)=\min_q\max(1-T_{tq},B_q(w)).
\end{equation}
For a letter \(a\), a bounded inclusion entry reduces
\(\max(1-\widetilde\ell_{t'}(w),\widetilde\ell_t(aw))\) over tested suffixes.
The initial entry similarly tests inclusion into the initial language table.
Masks exclude invalid source states and output slots. Supplementary separator
words refine transition entries; the current implementation does not apply
those refinements to the initial vector.

For exact compilation, the host constructs a DFA over positive formulas,
computes language-inclusion relations, and evaluates the selected
configuration languages against the initial and successor formulas. A
construction-size cap can make this procedure fail; it does not justify
silently treating an unresolved inclusion as true. Exported slots are ordered
canonically by state bitmask. The Boolean arrays are then sufficient for word
acceptance and carrier progression without revisiting the source graph.

The export contract checks encoding agreement, threshold consistency,
word acceptance against the operator audit, prefix-by-prefix stepping,
batched/single agreement up to slot permutation, and completion-distance
behavior. These checks validate the compiled object and its interface; they
do not replace equivalence checks against the task's language.

\mypara{Implementation scope.}
The experimental architecture uses both the encoder library and an
experimental pinned-state/formula/closure implementation. In particular,
validated library configuration fields alone do not establish that a
library-native checkpoint executes this adopted architecture. Source and
configuration fingerprints must accompany final checkpoints.
The released reproducibility archive includes the experimental launcher
options and checkpoint specifications alongside the library code. It records
which head and caps produced each checkpoint, including differences between
experimental environment options and library defaults.

\subsection{Feature arrays and pooling definitions}

Enumerate \(\Sigma=\{a_1,\ldots,a_m\}\), let \(n=\abs Q\), and
denote the hyperedge set by \(E\), with \(b=\abs E\). We assign features
\begin{equation}
  \begin{aligned}
    x_v&=\bigl(\mathbf{1}[v\in F],\mathbf{1}[v=v_\iota]\bigr)
      \in\Be^2,\\
    x_e&=\begin{cases}
      (\vect{e}_j,0), & e=(q,\tau,a_j),\\
      (\vect{0}_m,1), & e=(v_\iota,\tau,\mathrm{init}),
    \end{cases}
    \qquad x_e\in\Be^{m+1},
  \end{aligned}
  \label{eq:graph-features}
\end{equation}
where \(\vect{e}_j\in\Be^m\) is the \(j\)-th basis vector. Thus, the
vertex and edge feature matrices have shapes
\(X_V\in\Be^{(n+1)\times2}\) and \(X_E\in\Be^{b\times(m+1)}\).
Source and target incidences remain distinct relations.

Obligations and alternatives combine through coordinatewise smooth extremes,
\begin{equation}
  \smax_\epsilon(X)=\epsilon\log\sum_{x\in X}\exp(x/\epsilon),\qquad
  \smin_\epsilon(X)=-\smax_\epsilon(-X),
  \label{eq:encoder-extremes}
\end{equation}
with learned vectors for empty pools.
\subsection{Syntactic operator construction}

With the slot notation from the main paper, the syntactic construction
evaluates the assigned formulas on the selected configurations:
\begin{align}
  \alpha_t^{\mathrm{syn}}&=\mathbf{1}[\tau_t\models\widehat\iota],\nonumber\\
  M_a^{\mathrm{syn}}[t,t']&=
  \mathbf{1}\!\left[\tau_{t'}\models
    \bigwedge_{q\in\tau_t}\widehat\Delta(q,a)\right],
  \qquad \beta_t=\mathbf{1}[\tau_t\subseteq F].
  \label{eq:syntactic-operator}
\end{align}

\section{Training and Dataset Details}
\label{app:training}

\subsection{Corpora and source families}

The source generators provide finite-alphabet automata with both structured
and random language families. \cref{tab:families} records the two corpus
inventories available in the experiment documentation. Sequence, reach-avoid,
parallel-sequence, counting, and goal-set sources use conjunctions of ordered
letter chains. The \texttt{rad} generator is present in the ten-letter
comparison corpora and contributes no rows to these two inventories.

\begin{table}[htbp]
\centering
\caption{Counts of automaton presentations by source family in the four- and
sixteen-letter corpus inventories, before separating training and validation.
DFA/AFA denote deterministic/alternating finite automata. Counts are
presentations, not necessarily distinct languages; these inventories contain
no RAD-generated rows.}
\label{tab:families}
\begin{tabular}{lrr}
\toprule
Source & Four letters & Sixteen letters\\
\midrule
Random DFA & 316 & 143\\
Fresh random AFA & 719 & 146\\
Reverse random DFA & 832 & 398\\
\(k\)-th from end & 181 & 91\\
Modular counters & 108 & 70\\
Eventually conjunction & 100 & 89\\
Safety conjunction & 29 & 76\\
Locally testable & 320 & 151\\
Finite-trace pattern & 285 & 225\\
Sequence & 296 & 79\\
Reach-avoid & 346 & 130\\
Parallel sequences & 294 & 178\\
Counting & 126 & 84\\
Goal set & 62 & 188\\
RAD & 0 & 0\\
\midrule
Total & 4014 & 2048\\
\bottomrule
\end{tabular}
\end{table}

Rows have symbolic acceptance, residuals, and a stored configuration
automaton. A configuration-enumeration cap is recorded by a completeness flag;
truncated configuration data cannot certify complete reachable coverage.
The four-letter validation row classes are 149 plain, 119 conjunctive within
capacity, and 40 above capacity. The ten-letter split has 234 validation rows
(112 plain, 71 within-capacity product, 51 above capacity); the sixteen-letter
split has 159 (82, 59, 18).
Pretraining uses 1,396/1,043/684 presentations at four/ten/sixteen letters;
validation uses 308/234/159 and test uses 301/233/153. Every training
presentation has 2--8 source states (mean 5.13/5.12/4.91). Their minimal
DFAs can be larger: 468/298/228 training presentations have more than 16 DFA
states, with maxima 949/679/252. Thus small source presentations already
expose the encoder to languages of substantial deterministic complexity.
Splits hold out whole source families and nontrivial initial languages:
recomputed family and language-digest intersections are empty for each pair
of splits. Residual-language overlap is not excluded. Corpus histograms,
configuration hashes, and split-audit sources accompany the released data.

\subsection{Adopted settings}

\cref{tab:training-settings} records the settings supported by the
architecture and batch reports. It distinguishes the word-table setting
actually used by the four-letter semantic experiments from larger-depth
options. Checkpoint manifests and launcher options determine these settings.

\begin{table}[htbp]
\centering
\caption{Pretraining settings for the four-letter learned-assignment model with
semantic transitions (checkpoint family \texttt{c3lsem}). Steps count encoder
optimization updates; loss weights multiply the named supervision terms.
The copied-formula variant's loss differences are given in the final row.
Exact inclusion is used at export; training uses bounded semantic probes.}
\label{tab:training-settings}
\begin{tabular}{lp{.61\linewidth}}
\toprule
Setting & Value or specification\\
\midrule
Message passing & AND-OR, eight rounds, width 32.\\
Task readout & Initial-vertex state, width 32, without final normalization.\\
Source-state capacity & 16 pinned source-state vectors.\\
Operator capacity & 16 configuration slots.\\
Formula source & Copied conjuncts, learned disjunct assignments, copied acceptance.\\
Configuration source & Forward closure from thresholded initial terms.\\
Candidate cap & 64.\\
Operator & Semantic stepping; exact inclusion at evaluation/export.\\
Operator smoothing & Temperature 0.03.\\
Identity & Task-vector block only; weight 0.1.\\
Transition target & Weight 1 in the learned-assignment regime.\\
Step and closure losses & Off.\\
Training steps & 75,000 encoder optimization updates.\\
Semantic word-table depth & Words through length four; deeper tables are separate configurations.\\
Extra words & Up to 16 closure witnesses and 32 separators per graph.\\
Vertex-word target & Every word of length at most three; weight 0.3.\\
Word losses & Membership 1, prefix 1, witness 0.25.\\
Optimizer & AdamW, weight decay zero, global-norm clip 1.\\
Schedule & 500-step warmup to \(3\times10^{-4}\), cosine decay to \(10^{-5}\).\\
Copied regime & Identity weight zero; formula target weight 1, already satisfied.\\
\bottomrule
\end{tabular}
\end{table}

The source-state and formula caps constrain what the network can read, while
the sufficient-capacity corollary concerns what an operator can represent.
Inputs exceeding a source-state or edge cap require separate eligibility or
truncation accounting. Such limits must not be folded into a claim of
exactness for every AFA with a small configuration automaton.

The copied head has 12 disjunct positions and eight conjunct positions;
the learned-assignment head has eight and six. Identity supervision acts on
\(z\) with weights zero and 0.1, respectively. Both regimes retain the
per-vertex word head. The documented source-state cap is 16; the pinned
construction retains the first 16 input states when that cap is exceeded.
This input restriction is distinct from the operator's configuration budget.

\begin{table}[htbp]
\centering
\caption{Frozen encoder checkpoints for the representation and control experiments.
Alphabet size counts event symbols; each cell gives checkpoint ID and
pretraining updates (k means thousands). Suffix s identifies the encoder seed.
Policy-seed replication does not replicate encoder training. The corrected
LetterWorld comparison uses learned-assignment residual embeddings; it has
no copied-formula arm.}
\label{tab:encoder-bank-checkpoints}
\small
\begin{tabular}{lll}
\toprule
Alphabet & Copied formulas & Learned assignments\\
\midrule
4 & \texttt{c3c-s4}, 75k & \texttt{c3lsem-s0}, 75k\\
10 & \texttt{c3c10-s0--s2}, 75k & \texttt{c3lsem10-s0}, 75k\\
16 & \texttt{c3c16-s0}, 40k & \texttt{c3lsem16-short-s0}, 45k\\
\bottomrule
\end{tabular}
\end{table}

\subsection{Targets and sampling}
\label{app:sampling}

Membership batches include original and residual automata with exact word
labels, grouped by length and label. Prefix examples supervise every prefix
of a sampled long word; the documented prefix horizon is 64. Witness examples
include shortest distinguishing words and samples from a pair's symmetric
difference. The witness separation penalty is the sum of the two pointwise
membership losses and a squared hinge on their score difference.
Identity examples compare residual languages, with separate averages over
equal and different pairs. Transition targets evaluate decoded formulas on
input monomials, singleton state sets, the entire state set, and the empty set.
For copied formulas and accepting indicators, the input-derived probe targets
are identically satisfied. The configured formula weight remains one.

The closure selector's enumeration is a host-side discrete operation on
stopped-gradient factors. Its scoring model receives a straight-through
top-selection gradient. Formula factors also receive their direct transition
supervision; the existence of a closure candidate is not differentiated as
an exhaustive search tree.

\subsection{Semantic surrogate limitations}
\label{app:surrogate}

Bounded membership tables test inclusion on finitely many suffixes. A
separator is a word accepted by a proposed destination language that is not
in the source language's letter quotient. It invalidates a spurious
transition inclusion. A closure witness instead exposes a missing part of
a quotient not covered by available destination languages. These words have
different purposes even though both augment the bounded table.

The recorded ten-letter check on 30 rows obtains agreement of transition
inclusions with the exact test when separator refinement is uncapped. At a
cap of 32 separators, six rows retain inexact entries. The initial vector,
which is not refined by those extra words, differs on five rows. Exact
export recomputes both initial and transition inclusions and therefore has
a different fidelity contract from this training surrogate.
The reproducibility archive records per-run configuration/specification
SHA-256 hashes, launcher options, checkpoint paths, and source-file hashes.
The ten-letter copied runs have 75k checkpoints for seeds 0--2; seeds 3--4
are incomplete in the archive and are not counted as completed 75k runs.
Semantic word-table depths are 4/3/2 for four/ten/sixteen letters.
The four-letter learned model uses 32 separators and 16 closure witnesses;
the ten-letter model uses 192 and 16. Encoder seeds and policy seeds are
recorded separately in the manifests.

\subsection{Detailed supervision objectives}
\label{app:objectives}

Encoder and decoder pretraining uses
automata, residuals obtained by symbolic progression, and words with exact
acceptance labels. These labels let us supervise both the language behavior
of the operator and the semantic relationships between task vectors.
During training, we use soft arrays
\((\widetilde{\vect{\alpha}},\widetilde M,
\widetilde{\vect{\beta}})\) and compute word scores
\(\widetilde f_\Ac(w)\) in the max-min algebra on \([0,1]\).
On \(\{0,1\}\), these operations agree with Boolean acceptance.

\mypara{Acceptance and residual progression.}
For a word score \(s\), exact label \(y\in\{0,1\}\), threshold \(b\),
and margin \(m_{\mathrm{mem}}\), we define the pointwise membership loss
\begin{equation}
  \ell_{\mathrm{mem}}(s,y)=
  y[b+m_{\mathrm{mem}}-s]_+^2+
  (1-y)[s-b+m_{\mathrm{mem}}]_+^2,
  \label{eq:app-membership-loss}
\end{equation}
where \([x]_+=\max(x,0)\). The main paper's \(\operatorname{BalAvg}\)
first averages valid entries within each label/length-bucket group, then
present labels within a bucket, then present buckets. Membership uses
sampled continuation-length buckets; prefixes use geometric prefix-length
buckets and include \(\varepsilon\). To supervise progression, we score prefixes of sampled
words and also encode residual automata \(\Ac_u\), using the identity
\(\mathbf{1}[uv\in L(\Ac)]=\mathbf{1}[v\in L(\Ac_u)]\).
For language-distinct pairs, we use exact distinguishing words to supervise
both operators' labels and require the accepting score to exceed the
rejecting score by a margin. We denote these membership, prefix, and witness
losses by \(\Lc_{\mathrm{mem}}\), \(\Lc_{\mathrm{pre}}\), and
\(\Lc_{\mathrm{wit}}\), respectively.

\mypara{Semantic agreement of task vectors.}
For automata \(\Ac,\Ac'\) with languages \(L,L'\) and task vectors
\(z,z'\), we let \(d_z=\norm{z-z'}_2\). When their languages differ,
we let \(j\) denote the length of a shortest distinguishing word.
The identity loss is
\begin{equation}
  \ell_{\mathrm{id}}(z,z')=
  \begin{cases}
    d_z^2, & L=L',\\
    [m_{\mathrm{id}}\rho^j-d_z]_+^2, & L\ne L',
  \end{cases}
  \qquad 0<\rho\leq1.
  \label{eq:app-identity-loss}
\end{equation}
The sum of the separate equal- and unequal-language group means gives
\(\Lc_{\mathrm{id}}\). For the unequal-language term, the implementation
uses \(\sqrt{\max(d_z^2,10^{-12})}\) to keep the distance gradient finite;
the equal-language term uses the unfloored squared distance. This loss encourages equivalent
presentations to agree and distinct languages to separate, with a margin
that depends on the distinguishing-word length. We apply it to
\(z=h_{v_\iota}^{(L)}\), the task-vector component of the encoder output.

\mypara{State-language supervision.}
For a logit \(a\) and binary target \(y\),
\(\operatorname{BCEWithLogits}(a,y)=\log(1+\exp a)-ya\). Thus
\begin{equation}
 \Lc_{\mathrm{vertex}}=\operatorname{Avg}_{(v,w)}
 \left[\log(1+\exp a_{v,w})-\mathbf1[w\in L_v]a_{v,w}\right],
 \qquad w\in\mathcal W.
 \label{eq:vertex-loss}
\end{equation}

An auxiliary linear word head supplies supervision to the final vertex vectors.
For each word \(w\) in a bounded word table, the head predicts membership
in \(L_\Ac(q)\) at source vertex \(q\) and in \(L(\Ac)\) at the initial
vertex. We average binary cross-entropy against these exact labels to obtain
\(\Lc_{\mathrm{vertex}}\), giving the message-passing layers a direct
state-language target.

\mypara{Formula supervision and combined objective.}
In the learned-assignment regime, we supervise the decoded positive formulas
on probe sets of states. We label a probe set positive exactly when it
contains an input monomial, and apply the same construction to initial
formulas. This gives a formula loss \(\Lc_{\mathrm{tr}}\); in the
copied-formula regime, we supply the formulas directly.
For each of transition probes, initial probes, and accepting-state indicators,
the positive- and negative-label means are summed separately; the sum of
these three terms is \(\Lc_{\mathrm{tr}}\). Formula probes use the squared
hinge \(\ell_{.5,.2}\); acceptance targets are \(\mathbf1[q\in F]\).
The combined objective uses nonnegative weights:
\begin{equation}
  \begin{aligned}
    \Lc={}&\lambda_{\mathrm{mem}}\Lc_{\mathrm{mem}}
      +\lambda_{\mathrm{pre}}\Lc_{\mathrm{pre}}
      +\lambda_{\mathrm{wit}}\Lc_{\mathrm{wit}}\\
      &+\lambda_{\mathrm{vertex}}\Lc_{\mathrm{vertex}}
      +\lambda_{\mathrm{id}}\Lc_{\mathrm{id}}
      +\lambda_{\mathrm{tr}}\Lc_{\mathrm{tr}}.
  \end{aligned}
  \label{eq:app-training-objective}
\end{equation}
Copied formulas and accepting indicators already satisfy their probe
targets, so their formula penalty is zero. The operator losses also
supervise configuration selection through the selected candidates'
contribution to word acceptance.

\mypara{Differentiable inclusion.}
To train semantic stepping, we approximate the initial and successor language
inclusions defined in the main paper through state membership tables on
bounded-length words. We augment these tables with witnesses of uncovered
quotient languages and words that refute proposed inclusions.
Soft-array construction uses smooth reductions within \([0,1]\); word
evaluation uses hard max/min forward values with a soft gradient estimator. These finite tests provide a differentiable surrogate for language
inclusion. We use exact inclusion to construct the Boolean operator from
the selected configurations; its language fidelity also depends on which
configuration languages they cover, as we discuss next.

\section{Additional Representation Experiments}
\label{app:experiments}

\subsection{Evaluation definitions}

\mypara{Operator fidelity.}
Let \(N_6^{\mathrm{dec}}\) count rows whose compiled operator agrees with the
true language on every word of length at most six. For syntactic ablations,
\(N_6^{\mathrm{true}}\) repeats this test with true formulas and the selected
configurations, separating formula and configuration errors. These are the
\(D/B\) diagnostics in the source reports. All-word equivalence is a separate
check against the minimal DFA; bounded-word agreement alone does not establish
it. An empty-carrier false-dead rate counts prefixes at which the carrier is
empty while the true residual still has an accepting continuation. The
progress audit also reports a completion-distance false-dead rate, which
includes nonempty carriers without an accepting continuation inside its
search horizon. We keep these measurements separate.

\mypara{Latent fidelity.}
Pair AUC is the probability that two equal-language presentations have smaller
latent distance than a different-language pair, with the gallery protocol
held fixed. The fair gallery uses non-bisimilar input presentations. AFA and
minimal-DFA gallery comparisons are reported separately. A seed with AFA pair
AUC below \(0.6\) is recorded as collapsed under the experiment protocol.

\subsection{Why AND-OR message passing?}

The architecture study tests the local language recursion independently of
the final configuration decoder. Applying a hand-built Boolean AND-OR update
to length-\(K\) state language tables reproduces length-\(K+L\) tables after
\(L\) rounds, for \(K,L\in\{1,2,3\}\). The check records zero incorrect bits
across all 305 validation rows, every source state, and the initial expression;
the largest tested cell contains 10.4 million bits. The min/max pools implement
the recursive conjunction and disjunction directly.

A learned table-extension comparison uses width 32, 3,000 Adam steps,
and two seeds (\cref{tab:andor-local}). It measures newly predicted
word-table bits at the state and initial-vertex interfaces independently
of the configuration decoder.

\begin{table}[htbp]
\centering
\caption{Learned language-table extension, width 32 and 3,000 Adam steps.
Entries are validation accuracies on newly predicted membership bits for
seeds 0/1. \(2\to L\) extends words through length two to length \(L\);
States/Initial identify source-state/initial-expression readouts.
AND-OR and Sum denote the aggregation rules.}
\label{tab:andor-local}
\small
\begin{tabular}{lrrrr}
\toprule
& \multicolumn{4}{c}{Membership accuracy \(\uparrow\)}\\
\cmidrule(l){2-5}
Layer & States, \(2\to3\) & Initial, \(2\to3\) & States, \(2\to4\) & Initial, \(2\to4\)\\
\midrule
AND-OR & .949/.951 & .882/.885 & .902/.903 & .841/.847\\
Sum & .897/.898 & .843/.839 & .867/.868 & .810/.816\\
\bottomrule
\end{tabular}
\end{table}

The full-encoder comparison uses 75k training steps, three seeds, and
per-vertex language supervision. \cref{tab:andor-encoder} separates
initial-vertex and pooled-latent readouts in a configuration without pinned
slots or closure selection. The paired validation long-word error difference
is \(+0.013\;[-0.009,+0.034]\) for AND-OR minus sum.
The final architecture reads the task root and pinned vertex vectors directly;
this comparison does not isolate aggregation in that model.

\begin{table}[t]
\centering
\caption{Aggregation comparison after 75,000 training steps with per-vertex
language supervision. Entries give validation membership accuracies for
seeds 0/1/2. Initial and Pooled identify initial-vertex and pooled-task
readouts; lengths count events. Random DFAs use words of lengths one to
three. This configuration has no pinned slots or closure selection.}
\label{tab:andor-encoder}
\small
\setlength{\tabcolsep}{4pt}
\begin{tabular}{lrrrr}
\toprule
& \multicolumn{4}{c}{Membership accuracy \(\uparrow\)}\\
\cmidrule(l){2-5}
Layer & Initial, length 2 & Initial, length 3 & Random DFAs & Pooled, length 2\\
\midrule
AND-OR & .900/.897/.888 & .847/.848/.840 & .798/.823/.813 & .775/.772/.772\\
Sum & .820/.842/.860 & .787/.788/.793 & .709/.702/.710 & .787/.780/.783\\
\bottomrule
\end{tabular}
\end{table}
The local table-extension results support the algebraic aggregation rule.
The full-encoder comparison shows that this local advantage does not establish
a long-word advantage for the pooled readout; the final configuration pipeline
uses a different interface.

\subsection{Capacity and architectural ablations}

The budget/DFA-size partition has 210 rows with \(C\leq16\) and
\(\abs{D_\Ac}\leq16\), 58 with \(C\leq16\) and \(\abs{D_\Ac}>16\), and
40 with \(C>16\). The second group's median DFA size is 32 and its maximum
is 144.
These 58 rows demonstrate the distinction between a configuration budget
and a minimal-DFA state budget. A 16-slot operator can represent their
languages when coverage holds, even though their minimal DFAs exceed 16 states.

\begin{table}[t]
\centering
\caption{Four-letter validation after 75,000 training steps, with 16 operator
slots. Formula source, configuration selection, and transition construction
identify each model. Counts give rows agreeing with the true language on
all words through length six, among 119 within-capacity conjunctive products,
149 plain rows, and 40 rows with configuration count \(C>16\).
Ranges span the listed encoder seeds. FD is the fraction of product prefixes
with an empty carrier despite an accepting true continuation; these checks
do not establish all-word equivalence. A dash denotes an unavailable
bounded-word count.}
\label{tab:architecture}
\small
\begin{tabular}{llrrrrr}
\toprule
Formulas & \shortstack{Configurations /\\transitions} & Seeds & \shortstack{Product\\/119 \(\uparrow\)} & \shortstack{Plain\\/149 \(\uparrow\)} & \shortstack{\(C>16\)\\/40 \(\uparrow\)} & FD \(\downarrow\)\\
\midrule
Learned & per slot / syntactic & 5 & 43--53 & 130--144 & 2--4 & .046--.270\\
Learned & closure / syntactic & 5 & 95--100 & 120--132 & 2--4 & .031--.045\\
Learned & closure / semantic & 3 & 114--115 & 139--143 & 4--6 & .019\\
Copied & closure / syntactic & 5 & 119 & 149 & -- & .000\\
Copied & closure / semantic & 3 & 119 & 149 & 5--6 & .000\\
\bottomrule
\end{tabular}
\end{table}

The architectural question is which structural choices produce a faithful
executable task representation. The \texttt{lconjid}/\texttt{c3lid} comparison
examines per-slot configuration prediction versus selection from closure.
A frozen-checkpoint intervention replaces syntactic stepping by exact semantic
stepping while keeping the trained encoder and selected configurations fixed
(\cref{tab:restep}). Comparing that intervention with \texttt{c3lsem}
examines semantic construction during training as well as evaluation.
\texttt{c3csem}/\texttt{c3lsem} compares copied and learned formula-assignment
regimes with semantic stepping. A complete checkpoint-configuration comparison
is required before attributing a trained-arm difference to a single component.
The copied per-slot \texttt{cmaxstepid} arm is reported below with its
additional configuration differences identified.

\begin{table}[t]
\centering
\caption{Replacing syntactic transitions by exact semantic transitions while
holding each encoder and its configurations fixed. Counts give agreement
on every word through length six among 119 within-capacity conjunctive
validation rows. Ranges span the listed encoder seeds; gains are computed
within each checkpoint. Formula source and configuration selection identify
the models; other training differences remain uncontrolled.}
\label{tab:restep}
\begin{tabular}{lrrrr}
\toprule
Formulas / configurations & Seeds & Syntactic \(\uparrow\) & Semantic re-step \(\uparrow\) & Gain \(\uparrow\)\\
\midrule
Learned / per slot & 0--3 & 44--53 & 63--67 & 10--22\\
Copied / per slot & 0--1 & 52--59 & 73--75 & 16--21\\
Learned / closure & 0--2 & 98--100 & 105--107 & 5--9\\
\bottomrule
\end{tabular}
\end{table}
Exact semantic re-stepping improves bounded-word agreement while holding
the learned features and selected configurations fixed. This intervention
provides direct evidence that algebraic language relations improve execution
beyond the structural representation alone. Comparisons between separately
trained models also include training and configuration-selection differences.

\subsection{Semantic agreement and progress}

\begin{table}[t]
\centering
\caption{Task-vector agreement on non-bisimilar validation presentations.
Pair AUC measures whether equal-language pairs are closer than different-language
pairs; larger is better. AFA and DFA columns use alternating-automaton and
minimal-DFA galleries. Ranges span encoder seeds; a seed is counted as
collapsed when AFA AUC is below 0.6. Learned/copied describe formula sources;
all models use syntactic transitions except Learned/semantic. Closure models
use candidate selection; no identity omits task-vector identity supervision.}
\label{tab:latent}
\small\setlength{\tabcolsep}{4pt}
\begin{tabular}{lrrr}
\toprule
Model & AFA pair AUC \(\uparrow\) & DFA pair AUC \(\uparrow\) & Collapsed / seeds \(\downarrow\)\\
\midrule
Learned / per slot & .914--.960 & .774--.815 & 0/5\\
Learned / closure, no identity & .504--.888 & .797--.936 & 1/5\\
Learned / closure & .954--.967 & .758--.872 & 0/5\\
Learned / semantic & .944--.962 & .801--.860 & 0/3\\
Copied / closure & .845--.922 & .896--.951 & 0/5\\
\bottomrule
\end{tabular}
\end{table}

The progress audit samples ten random prefixes per validation row, with
lengths zero through six, and includes prefixes of a shortest accepted word.
For \texttt{c3c-s4}, completion distances equal the true distances on all
sampled plain and product prefixes. For \texttt{c3lsem-s0}, equality shares
are \(0.943\) and \(0.980\), respectively; completion-distance false-dead
rates are \(0.057\) and \(0.020\). These are single-checkpoint audits with
search horizon 64. Further class-specific diagnostics appear below.
The pair-ranking and execution audits assess distinct properties. High pair
AUC does not establish exact equality across language-equivalent presentations,
and exact copied execution does not imply optimal learned feature geometry.

\cref{fig:tsne-sources} gives a qualitative view of the source-family
vectors. Its samples come from the source corpus, not exclusively the
validation split used in \cref{tab:latent}.
\begin{figure}[t]
\centering
  \includestandalone[]{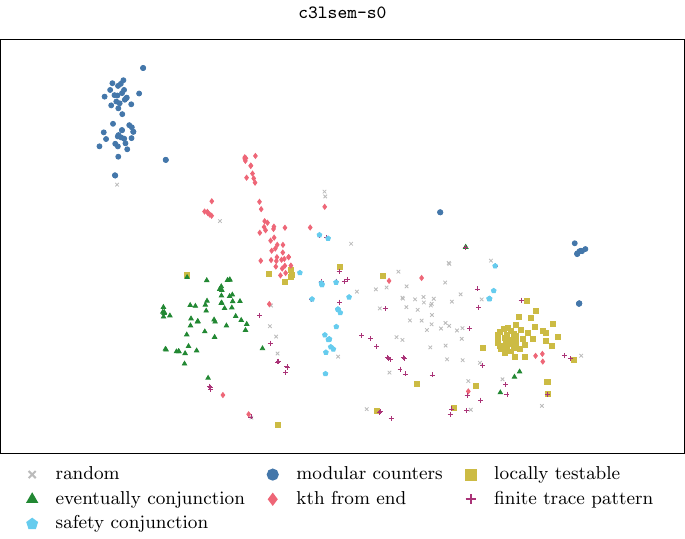}
\caption{Source-family task vectors from the learned-assignment, semantic-transition
encoder (\texttt{c3lsem-s0}), projected by t-SNE with perplexity 30 and seed 0.
Up to 60 rows per structured family and 60 total random-family rows (gray)
are sampled from the source corpus, including possible training rows.
Axes are projection coordinates; distances and cluster gaps are qualitative,
with semantic statistics measured in the original latent space.}
\label{fig:tsne-sources}
\end{figure}

\subsection{Per-seed representation results}

\cref{tab:per-seed} records the adopted copied and learned closure
configurations. Counts are bounded-word row agreements, not a conversion of
the entire validation table into all-word equivalence statements. The copied
syntactic checkpoints are from batch J and the semantic checkpoints from
batch K. Distinct batches are not assumed to be identical realizations merely
because their seed indices agree.

\begin{table}[htbp]
\centering
\caption{Four-letter validation per encoder seed after 75,000 steps.
Models use closure-selected configurations and the indicated formula source
and transitions. Product/plain/above-capacity counts measure agreement on
all words through length six, with denominators 119/149/40 and capacity 16.
FD counts empty-carrier product prefixes with a live true residual. AFA/DFA
pair AUC ranks equal-language pairs ahead of different-language pairs in
alternating/minimal-DFA galleries. Dashes denote unavailable bounded-word counts.}
\label{tab:per-seed}
\small
\begin{tabular}{lrrrrrr}
\toprule
Model, seed & Product \(\uparrow\) & Plain \(\uparrow\) & \shortstack{Above\\capacity \(\uparrow\)} & FD \(\downarrow\) & \shortstack{AFA AUC\\\(\uparrow\)} & \shortstack{DFA AUC\\\(\uparrow\)}\\
\midrule
Copied/syntactic, 0 & 119 & 149 & -- & .000 & .894 & .950\\
Copied/syntactic, 1 & 119 & 149 & -- & .000 & .845 & .896\\
Copied/syntactic, 2 & 119 & 149 & -- & .000 & .913 & .947\\
Copied/syntactic, 3 & 119 & 149 & -- & .000 & .865 & .938\\
Copied/syntactic, 4 & 119 & 149 & -- & .000 & .922 & .951\\
Copied/semantic, 0 & 119 & 149 & 5 & .000 & .945 & .965\\
Copied/semantic, 1 & 119 & 149 & 6 & .000 & .879 & .930\\
Copied/semantic, 2 & 119 & 149 & 5 & .000 & .901 & .943\\
Learned/semantic, 0 & 115 & 139 & 6 & .019 & .962 & .815\\
Learned/semantic, 1 & 114 & 141 & 4 & .019 & .951 & .860\\
Learned/semantic, 2 & 115 & 143 & 4 & .019 & .944 & .801\\
\bottomrule
\end{tabular}
\end{table}
For the copied syntactic model, a separate all-word audit of the thresholded
operator gives above-capacity exact-language counts of 3, 5, 4, 4, and 3
out of 40 for seeds 0--4. These counts use a different metric from the
bounded-word columns and are reported separately.

\subsection{Architecture and supervision controls}

The copied per-slot \texttt{cmaxstepid} configuration has five seeds. Its
product counts are \(52,59,52,53,55\); its plain counts are
\(148,149,147,148,149\). It has four above-capacity rows in every seed.
Its aggregation and supervision
settings must be retained when comparing it with closure arms.

Identity supervision is a separate intervention: \texttt{c3l}/\texttt{c3lid}
compares the closure configuration without/with task-vector identity loss;
its collapsed-seed counts are one of five and zero of five.
The identity comparison records fewer collapsed seeds with task-vector
supervision. It does not isolate every configuration difference between the
per-slot, closure, and semantic model families.

For \texttt{c3lsem} against \texttt{c3lid}, the validation long-word excess
difference is \(-0.024\), with interval \([-0.041,-0.010]\), denoting
fewer errors.

\subsection{Wider alphabets and progress audits}

At ten letters, the 75k-step \texttt{c3lsem10-s0} checkpoint has
all-word operator equivalence on 104/112 plain, 66/71 within-capacity product,
and 6/51 above-capacity validation rows. These are exact-language checks of
the thresholded operator. Its composition audit gives exact conjunction and
disjunction on 29/30 and 24/30 sampled pairs. The corrected LetterWorld
comparison below evaluates residual embeddings from this checkpoint.

For completion-distance audits, 10 random prefixes of length 0--6 and the
prefixes of a shortest accepted word are evaluated per row. A failed distance
search is reported separately from an empty carrier. For the adopted learned
checkpoint, the plain/product/above-capacity equal-distance shares are
.943/.980/.643, and the corresponding distance false-dead shares are
.057/.020/.348. The copied \texttt{c3c-s4} shares are
1.000/1.000/.670 and .000/.000/.328. False-alive distance shares are zero on
these sampled prefixes. The horizon is 64 events.
Copied execution preserves completion distances on the sampled within-budget
prefixes. Learned execution retains high agreement there, with more errors
on above-capacity rows. These finite-prefix audits use the stated search
horizon and do not establish distances on every possible prefix.

\subsection{Composition measurements and visualizations}

The main composition table uses \texttt{c3c-s4} for copied formulas with
syntactic transitions, \texttt{c3lsem-s0} for learned assignments with
semantic transitions, and \texttt{c3lid-s0} for learned assignments with
syntactic transitions.

\ifextendedpaper\else
\begin{figure}[t]
\centering
  \includestandalone[]{figures/standalone/tsne-composition}
\caption{t-SNE of validation components and compositions. Panels show learned
assignments/semantic transitions (\texttt{c3lsem-s0}), copied formulas/syntactic
transitions (\texttt{c3c-s4}), and learned assignments/syntactic transitions
(\texttt{c3lid-s0}). Gray triangles are components; circles/squares are
conjunctions/disjunctions, and crosses are conjunctions above the product
bound. \(C_i\) counts component configurations; operator capacity is 16.
Hollow markers replace a component by a language-equivalent presentation.
Axes are qualitative projection coordinates, not language distances.}
\label{fig:tsne-composition}
\end{figure}

\fi

\begin{figure}[p]
\centering
  \includestandalone[]{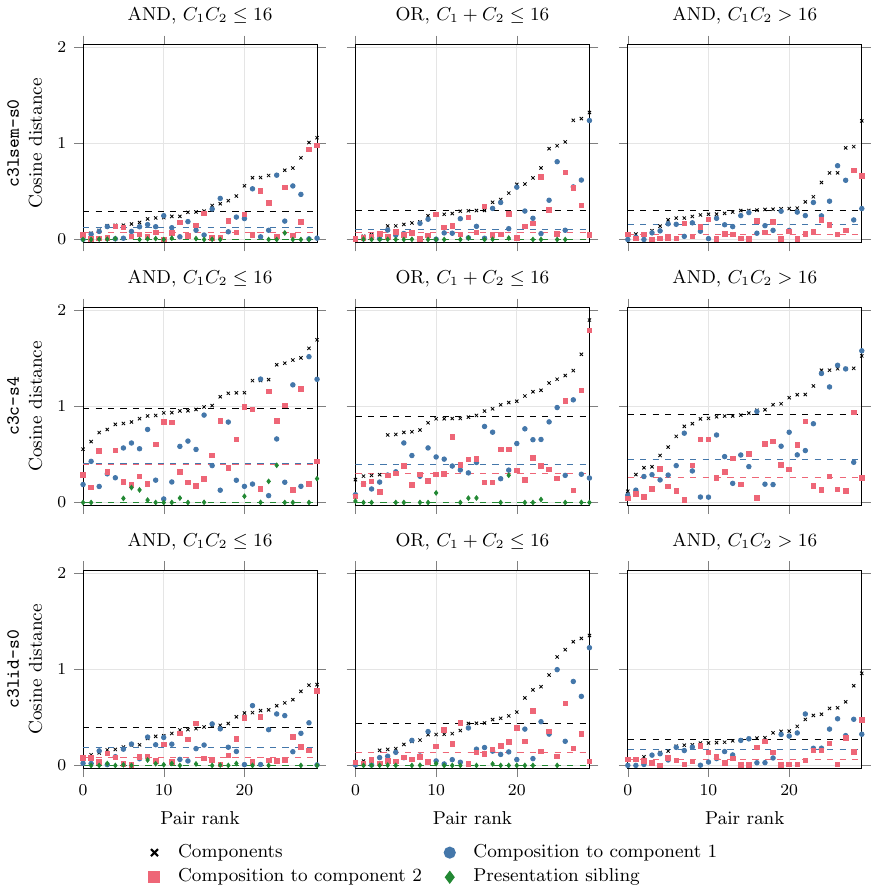}
\caption{Cosine distances for 30 validation pairs per panel. Rows show learned
assignments/semantic transitions (\texttt{c3lsem-s0}), copied formulas/syntactic
transitions (\texttt{c3c-s4}), and learned assignments/syntactic transitions
(\texttt{c3lid-s0}). Columns give conjunction with \(C_1C_2\leq16\),
disjunction with \(C_1+C_2\leq16\), and conjunction with \(C_1C_2>16\);
\(C_i\) counts component configurations. Horizontal position ranks pairs
by component-to-component distance (crosses). Circles/squares compare the
composition to each component; diamonds compare language-equivalent sibling
compositions where available. Dashed lines are medians. Product bounds need
not equal actual composed configuration counts.}
\label{fig:composition-distances}
\end{figure}

The main composition table uses all-word equivalence audits. A shortest-counterexample search that
hits a resource cap must be reported unresolved; the reported composition
audit records no unresolved pairs for these checkpoints.

Replacing a component by a language-equivalent presentation sibling provides
a separate test of composition latents. For \texttt{c3lsem-s0}, median cosine
distance to the sibling composition is .000 for conjunction and disjunction,
with maxima .069 and .016. For \texttt{c3c-s4}, the medians are .002 and .000,
with maxima .387 and .283. For \texttt{c3lid-s0}, both medians are .000,
with maxima .057 and .019. These numbers use the available sibling pairs,
not every composed pair.

For language-pair distance ordering, let \(j(P)\) be the shortest
 distinguishing-word length of pair \(P\), and \(d(P)\) its latent distance.
The ordering AUC counts \(d(P)<d(Q)\) among \(j(P)>j(Q)\), with ties weighted
by one half. \cref{tab:distance-ordering} uses 25,994 language pairs
with finite witness length and a bootstrap over 229 items. The identity
objective already uses witness-length-dependent margins; this diagnostic
measures a fuller pairwise ordering than that objective specifies.

\begin{table}[htbp]
\centering
\caption{Cosine-distance ordering AUC on 25,994 language pairs with finite
shortest distinguishing-word length \(j\). AUC counts comparisons where
\(j(P)>j(Q)\) and latent distance \(d(P)<d(Q)\), with half credit for ties;
larger is better. The last column restricts \(j(P)-j(Q)\geq2\).
Parentheses give 95\% bootstrap intervals over 229 items. Row labels identify
formula source, transition construction, and encoder seed.}
\label{tab:distance-ordering}
\begin{tabular}{lrr}
\toprule
Model, seed & All witness-length gaps \(\uparrow\) & Gaps at least two \(\uparrow\)\\
\midrule
Learned/semantic, 0 & .645 (.612--.680) & .787 (.731--.832)\\
Copied/syntactic, 4 & .724 (.676--.773) & .895 (.862--.924)\\
Learned/syntactic, 0 & .665 (.634--.701) & .834 (.787--.878)\\
\bottomrule
\end{tabular}
\end{table}
Equivalent-presentation substitutions usually give small composition-vector
distances, but the nonzero maxima preclude exact learned canonicity. Distance
ordering measures an empirical association with distinguishing-word length;
it does not certify an isometric language embedding.

\subsection{Failure trace}

The \texttt{c3lid-s0} trace identifies 27 plain rows with at least one
completion-distance false-dead prefix: 24 also fail the bounded row test and
three pass that bounded test. Ten rows have an empty initial carrier, six
have nonempty carriers steered away from acceptance, five lose a missing
disjunct at a killing step, and seven die with complete decoded step
formulas. These categories overlap. No acceptance-bit errors were found in
that trace. The ten empty-start cases have disjunctive initial formulas
whose thresholded decoding is the empty conjunction, while their compiled
initial carriers are empty. The discrepancy involves the initial-formula
head and its soft/hard interpretation as well as closure seeding.
The overlapping failures involve formula decoding, initialization, and
progression. They therefore do not identify configuration selection as the
sole source of error, nor a representational inability to express disjunction.

\subsection{Measured evaluation scope}

\begin{table}[t]
\centering
\caption{Measured experimental scope. Capacities count source states,
operator slots, and closure candidates; DFA size counts minimal-DFA states.
Environment frames are physical steps. Unmeasured time and memory costs
are listed explicitly; varying alphabet or task-pool size does not measure
scaling with operator capacity.}
\label{tab:scaling-scope}
\small
\begin{tabular}{lp{.65\linewidth}}
\toprule
Axis & Recorded scope\\
\midrule
Alphabet & 4, 10, and 16 letters; configurations differ across alphabets.\\
Representation & 16 source-state slots, 16 operator slots, at most 64 candidates.\\
DFA size & Within-budget four-letter rows include minimal DFAs with 144 states.\\
Training tasks & XLand: 25 versus 87 RAD-compatible, 139 multi-event, or 173 plain tasks, at fixed frames.\\
Policy samples & XLand: 1,000 updates at \(128\times128\) steps (16.384M);
  LetterWorld: 611 updates at \(256\times128\) steps (20.021M).\\
Compilation cost & Exact inclusion and bank construction; time and peak memory unmeasured.\\
\bottomrule
\end{tabular}
\end{table}

\section{Reinforcement Learning Protocols and Results}
\label{app:rl}

\subsection{Tasks, environments, and banks}

Policy training uses frozen, pretrained encoders. Task banks store residual
embeddings computed offline, so residual encoding cost is paid before
these RL comparisons.

For the main XLand comparison, the RAD-compatible and multi-event pools contain plain and product
rows with shortest accepted words of length at least two.
The
\texttt{xl4-multi} pool permits minimal DFAs with at most 128 states;
\texttt{xl4-rad} restricts this to ten for the RAD comparison.
Thus
``multi'' denotes a multi-event requirement and does not require every task
to have conjunctive structure.
In the frozen selection files, the large
training, validation, and test pools are pairwise disjoint in their stored
language digests, structural digests, and family identifiers for both pools.
This check concerns policy-task splits. The plain pool contains 37 test tasks
and includes single-event objectives; its training pools contain 25 and 173
tasks. The multi-event and RAD-compatible training pools contain 25/139 and
25/87 tasks, respectively. These pool definitions should not be interpreted
as disjoint test distributions across pools.

The comparison treats a task as an event language and separates its monitor
from task-free physical dynamics.
LetterWorld \citep{vaezipoor2021ltl2action} uses discrete movement on
letter grids.
Zones \citep{vaezipoor2021ltl2action} uses a continuous-control point robot and four color
propositions.
The XLand variants adapt XLand-MiniGrid \citep{nikulin2023xlandminigrid}
and use discrete actions and environment event
labels.
Each environment adapter specifies observation/action dimensions,
time limits, event labels, and physical termination independently of the
selected task.
Rising-edge conversion and the order of simultaneous events
must be identical for true and operator run graphs.

Morphata and RAD banks retain the same ordered task identities, true-DFA
transitions, accepting states, event maps, and distances.
Morphata stores
embeddings of residual source AFAs; RAD stores embeddings of minimal DFAs
with residual initial states.
A Morphata operator bank additionally stores
\(\alpha,M,\beta\).
Reachable carriers are enumerated from \(\alpha\) and
used for distance lookup and disagreement diagnostics.
A carrier-count cap
causes an error rather than silently truncating the graph.

\subsection{Comparison settings}

Initial arms provide a 32-dimensional task vector fixed for the episode;
the task-blind control receives zeros.
Residual arms read a 32-dimensional
vector at the current true-DFA row.
Operator conditioning concatenates the
initial vector with 16 Boolean carrier coordinates, yielding width 48 for
the adopted model.
These different input widths must be recorded when
comparing policy capacity.
Carrier-vector conditioning (\texttt{m-opv}) concatenates the initial
32-dimensional latent with a 32-dimensional mean of active configuration-slot
vectors, giving width 64.
A slot vector is the mean of the pinned vectors
of its source-state configuration; empty configurations and empty active
sets give zero.
The raw carrier interface has width 48.
These are distinct
inputs to the same conditioning projection.

All arms use true-DFA reward and termination in the comparison.
Carrier
errors do not terminate a DFA-monitored episode.
Evaluation reports true success, rejection,
truncation, unshaped return, and episode length.
Physical termination can
also stop an episode and must remain a distinct outcome when it produces
no task verdict.

XLand uses 25 training tasks per pool or larger pools of 87 RAD-compatible,
139 multi-event, and 173 plain tasks.
Within a table, arms use
matched tasks, layouts, seeds, and training budget.
The split and bank manifests record the task selections and encoder identities
for each pool.
RAD eligibility is checkpoint-specific.
The published checkpoint has limits
of ten DFA states and ten active symbols.
We match comparisons to the capacities
and task selections of the weights actually used.
Bank provenance records event-symbol and token assignments.

\subsection{Training, selection, and reporting}
\label{app:rl-protocol}

The code and documentation will be released at
\url{https://git.anandb.dev/morphata.git} and
\url{https://docs.anandb.dev/morphata/}. The frozen figure exports record
source-manifest and metrics-file SHA-256 hashes. All corrected control runs
use implementation revision \texttt{ef27a9a3} (change \texttt{oxwqzzol}),
with JAX 0.11.1, Flax 0.12.10, and Optax 0.2.8. The reproducibility archive
includes the complete PPO \citep{schulman2017proximal} configurations, bank hashes, and individual
policy-seed test means. All bank hashes match the frozen manifests.

XLand uses a linear observation encoder, a shared three-layer torso of width
128, and concatenated conditioning. Each of 1,000 PPO updates collects 128
steps from 128 environments, totaling 16,384,000 transitions. Orthogonal
initialization and learning-rate decay are disabled. Training-time validation
runs every 50 updates with 512 episodes. The two rejection rewards, 0 and $-1$,
leave the acceptance reward ($+1$), nonterminal reward (0), and termination
semantics unchanged: both acceptance and rejection end an episode.

Both environments use Adam at constant learning rate $3\times10^{-4}$,
global gradient-norm clipping at .5, four PPO epochs, GAE parameter .95,
PPO clip .2, value-loss coefficient .5, and entropy coefficient .01.
XLand uses discount .99 and four minibatches of 4,096 transitions per epoch;
LetterWorld uses discount .94 and 128 minibatches of 256 transitions.
Reward shaping is disabled. Final test evaluation samples policy actions
for one episode per task and frozen layout key, after the last update.
There are eight layout keys (seed 10001 for XLand). XLand evaluation is
capped at 75 steps; LetterWorld requests 128 steps, but its environment's
75-step time limit binds.

The observation/view grid was inspected on test outcomes and is exploratory.
Subsequent architecture selection used validation success with 87 training tasks averaged
over the last five validation points and over Morphata/RAD residual arms.
The shared depth-three torso led the next-best candidate by .003. Reward and
optimizer selection then used the same temporal average over Morphata residual
and carrier-vector arms; reward 0 led the reference by .005. The reward-$-1$
panel was added after inspecting step-three test outcomes. These development
choices preclude describing the entire procedure as test-independent.
The RAD-compatible confirmation repeats development seed indices 0--4 for
several arms under nondeterministic GPU training. Its runs are reported
separately and are not pooled with development runs as extra replications.

Test intervals use 10,000 crossed bootstrap draws over policy seeds, tasks,
and layouts. Differences share task/layout draws and resample policy seeds
independently across arms. All five policy seeds are retained, including
failed-learning outcomes; the intervals condition on the frozen encoder.
Validation bands use 2,000 seed-only resamples at each update and therefore
measure a different uncertainty from the test intervals.

\subsection{Normalization}
For standardized banks, a per-coordinate mean $\mu$ and a single scale
$s=d^{-1}\sum_i\operatorname{sd}_i$ are computed from nonzero latent rows
reachable in the large training pool. Nonzero rows use $z'=(z-\mu)/s$;
zero rows remain zero. Nonzero slot vectors use their own mean and scalar
scale from that pool, applied before carrier pooling. These transforms do
not change the Boolean operator arrays.

\begin{table}[htbp]
\centering\small
\caption{XLand test results with 87/139/173 training tasks
(RAD-compatible/multi-event/plain) with 95\% crossed seed/task/layout
intervals. This expands the main table; all five policy seeds contribute.}
\label{tab:xland-large-full}
\begin{tabular*}{\linewidth}{@{\extracolsep{\fill}}lccc@{}}
\toprule
Conditioning & RAD-compatible & Multi-event & Plain\\
\midrule
\multicolumn{4}{l}{\emph{Rejection reward 0 (validation-selected)}}\\
Task-blind & 0.510\,{\scriptsize (0.304--0.713)} & 0.623\,{\scriptsize (0.471--0.768)} & 0.749\,{\scriptsize (0.620--0.868)}\\
Ours/Initial & 0.560\,{\scriptsize (0.354--0.759)} & 0.634\,{\scriptsize (0.499--0.761)} & 0.775\,{\scriptsize (0.648--0.886)}\\
Ours/Residual & 0.656\,{\scriptsize (0.444--0.850)} & 0.770\,{\scriptsize (0.637--0.888)} & 0.828\,{\scriptsize (0.709--0.927)}\\
Ours/Carrier bits & 0.551\,{\scriptsize (0.349--0.750)} & 0.662\,{\scriptsize (0.521--0.795)} & 0.759\,{\scriptsize (0.629--0.874)}\\
Ours/Carrier vectors & 0.653\,{\scriptsize (0.448--0.841)} & 0.773\,{\scriptsize (0.641--0.891)} & 0.833\,{\scriptsize (0.716--0.930)}\\
RAD/Initial & 0.516\,{\scriptsize (0.307--0.721)} & -- & --\\
RAD/Residual & 0.709\,{\scriptsize (0.519--0.872)} & -- & --\\
\midrule
\multicolumn{4}{l}{\emph{Rejection reward -1 (reference)}}\\
Task-blind & 0.307\,{\scriptsize (0.082--0.581)} & 0.240\,{\scriptsize (0.000--0.523)} & 0.758\,{\scriptsize (0.629--0.874)}\\
Ours/Initial & 0.429\,{\scriptsize (0.231--0.631)} & 0.505\,{\scriptsize (0.359--0.659)} & 0.771\,{\scriptsize (0.645--0.882)}\\
Ours/Residual & 0.647\,{\scriptsize (0.434--0.840)} & 0.770\,{\scriptsize (0.637--0.887)} & 0.821\,{\scriptsize (0.703--0.920)}\\
Ours/Carrier bits & 0.457\,{\scriptsize (0.259--0.659)} & 0.634\,{\scriptsize (0.487--0.771)} & 0.752\,{\scriptsize (0.622--0.869)}\\
Ours/Carrier vectors & 0.640\,{\scriptsize (0.429--0.834)} & 0.755\,{\scriptsize (0.623--0.877)} & 0.833\,{\scriptsize (0.715--0.932)}\\
RAD/Initial & 0.434\,{\scriptsize (0.226--0.641)} & -- & --\\
RAD/Residual & 0.700\,{\scriptsize (0.509--0.872)} & -- & --\\
\bottomrule
\end{tabular*}

\end{table}

\subsection{Training-pool and reward comparisons}
\cref{tab:xland-small} gives the small-training-pool outcomes.
Carrier vectors do not uniformly improve control: on the plain pool they
underperform task-blind inputs with 25 training tasks under both rewards. With the larger
training pools (main table), carrier vectors exceed carrier bits and task-blind
inputs in all three pools under both rewards. \cref{tab:xland-interface-differences,tab:xland-differences-small} report the corresponding differences.

\begin{table}[htbp]
\centering\small\setlength{\tabcolsep}{4pt}
\caption{XLand with 25 training tasks, 1,000 PPO updates, five policy seeds.
Mean test success and 95\% crossed-bootstrap intervals on the same test pools
and layouts as the main results table.}
\label{tab:xland-small}
\begin{tabular*}{\linewidth}{@{\extracolsep{\fill}}lccc@{}}
\toprule
Conditioning & RAD-compatible & Multi-event & Plain\\
\midrule
\multicolumn{4}{l}{\emph{Rejection reward 0 (validation-selected)}}\\
Task-blind & 0.532\,{\scriptsize (0.319--0.738)} & 0.602\,{\scriptsize (0.455--0.746)} & 0.739\,{\scriptsize (0.608--0.859)}\\
Ours/Initial & 0.526\,{\scriptsize (0.322--0.725)} & 0.578\,{\scriptsize (0.442--0.710)} & 0.755\,{\scriptsize (0.626--0.870)}\\
Ours/Residual & 0.585\,{\scriptsize (0.378--0.782)} & 0.700\,{\scriptsize (0.560--0.829)} & 0.742\,{\scriptsize (0.616--0.855)}\\
Ours/Carrier bits & 0.424\,{\scriptsize (0.237--0.619)} & 0.492\,{\scriptsize (0.352--0.635)} & 0.689\,{\scriptsize (0.555--0.813)}\\
Ours/Carrier vectors & 0.601\,{\scriptsize (0.399--0.791)} & 0.676\,{\scriptsize (0.542--0.803)} & 0.657\,{\scriptsize (0.520--0.787)}\\
RAD/Initial & 0.437\,{\scriptsize (0.228--0.650)} & -- & --\\
RAD/Residual & 0.606\,{\scriptsize (0.399--0.794)} & -- & --\\
\midrule
\multicolumn{4}{l}{\emph{Rejection reward -1 (reference)}}\\
Task-blind & 0.409\,{\scriptsize (0.179--0.663)} & 0.575\,{\scriptsize (0.427--0.722)} & 0.746\,{\scriptsize (0.614--0.866)}\\
Ours/Initial & 0.488\,{\scriptsize (0.291--0.688)} & 0.467\,{\scriptsize (0.313--0.626)} & 0.732\,{\scriptsize (0.601--0.851)}\\
Ours/Residual & 0.579\,{\scriptsize (0.372--0.777)} & 0.681\,{\scriptsize (0.543--0.815)} & 0.737\,{\scriptsize (0.615--0.851)}\\
Ours/Carrier bits & 0.468\,{\scriptsize (0.272--0.666)} & 0.501\,{\scriptsize (0.359--0.646)} & 0.678\,{\scriptsize (0.542--0.804)}\\
Ours/Carrier vectors & 0.582\,{\scriptsize (0.382--0.771)} & 0.667\,{\scriptsize (0.535--0.795)} & 0.656\,{\scriptsize (0.518--0.787)}\\
RAD/Initial & 0.393\,{\scriptsize (0.185--0.612)} & -- & --\\
RAD/Residual & 0.601\,{\scriptsize (0.394--0.797)} & -- & --\\
\bottomrule
\end{tabular*}

\end{table}

\begin{table}[htbp]
\centering\small\setlength{\tabcolsep}{4pt}
\caption{XLand success differences with 87/139/173 training tasks
(RAD-compatible/multi-event/plain), first arm minus second, with 95\%
crossed-bootstrap intervals. Values are success-rate differences, not percentage
points. All five policy seeds and eight layouts per task are included.}
\label{tab:xland-interface-differences}
\begin{tabular}{llcc}
\toprule
Pool & Comparison & Reward $0$ & Reward $-1$\\
\midrule
RAD-compatible & \texttt{m-opv} $-$ \texttt{none} & +0.143 (+0.031, +0.278) & +0.332 (+0.094, +0.609)\\
RAD-compatible & \texttt{m-opv} $-$ \texttt{m-op} & +0.101 (+0.015, +0.213) & +0.182 (+0.056, +0.338)\\
RAD-compatible & \texttt{m-opv} $-$ \texttt{m-res} & -0.003 (-0.066, +0.057) & -0.007 (-0.066, +0.047)\\
RAD-compatible & \texttt{r-res} $-$ \texttt{m-res} & +0.053 (-0.051, +0.171) & +0.053 (-0.051, +0.166)\\
\addlinespace
Multi-event & \texttt{m-opv} $-$ \texttt{none} & +0.150 (+0.052, +0.262) & +0.515 (+0.239, +0.791)\\
Multi-event & \texttt{m-opv} $-$ \texttt{m-op} & +0.110 (+0.030, +0.205) & +0.121 (+0.034, +0.223)\\
Multi-event & \texttt{m-opv} $-$ \texttt{m-res} & +0.003 (-0.050, +0.057) & -0.015 (-0.076, +0.043)\\
\addlinespace
Plain & \texttt{m-opv} $-$ \texttt{none} & +0.084 (+0.016, +0.164) & +0.075 (+0.011, +0.151)\\
Plain & \texttt{m-opv} $-$ \texttt{m-op} & +0.074 (+0.012, +0.149) & +0.081 (+0.018, +0.157)\\
Plain & \texttt{m-opv} $-$ \texttt{m-res} & +0.005 (-0.034, +0.041) & +0.012 (-0.024, +0.049)\\
\addlinespace
\bottomrule
\end{tabular}

\end{table}
\begin{table}[htbp]
\centering\small\setlength{\tabcolsep}{4pt}
\caption{XLand success differences with 25 training tasks under the same evaluation and
bootstrap protocol as \cref{tab:xland-interface-differences}.}
\label{tab:xland-differences-small}
\begin{tabular}{llcc}
\toprule
Pool & Comparison & Reward $0$ & Reward $-1$\\
\midrule
RAD-compatible & \texttt{m-opv} $-$ \texttt{none} & +0.069 (-0.043, +0.201) & +0.174 (-0.015, +0.404)\\
RAD-compatible & \texttt{m-opv} $-$ \texttt{m-op} & +0.178 (+0.048, +0.332) & +0.115 (-0.015, +0.249)\\
RAD-compatible & \texttt{m-opv} $-$ \texttt{m-res} & +0.016 (-0.068, +0.116) & +0.003 (-0.093, +0.112)\\
RAD-compatible & \texttt{r-res} $-$ \texttt{m-res} & +0.021 (-0.078, +0.129) & +0.022 (-0.094, +0.156)\\
\addlinespace
Multi-event & \texttt{m-opv} $-$ \texttt{none} & +0.074 (-0.034, +0.189) & +0.092 (-0.022, +0.209)\\
Multi-event & \texttt{m-opv} $-$ \texttt{m-op} & +0.184 (+0.091, +0.287) & +0.166 (+0.076, +0.265)\\
Multi-event & \texttt{m-opv} $-$ \texttt{m-res} & -0.024 (-0.091, +0.045) & -0.014 (-0.072, +0.045)\\
\addlinespace
Plain & \texttt{m-opv} $-$ \texttt{none} & -0.082 (-0.180, -0.003) & -0.090 (-0.199, -0.003)\\
Plain & \texttt{m-opv} $-$ \texttt{m-op} & -0.032 (-0.124, +0.047) & -0.022 (-0.116, +0.057)\\
Plain & \texttt{m-opv} $-$ \texttt{m-res} & -0.084 (-0.178, -0.007) & -0.081 (-0.170, -0.005)\\
\addlinespace
\bottomrule
\end{tabular}

\end{table}

The reward response differs across interfaces. With 87 training tasks in the RAD-compatible
pool, removing the rejection penalty changes task-blind success from .307 to
.510, timeout share from .662 to .418, rejection share from .031 to .072, and
mean episode length from 56.4 to 42.3. Residual and carrier-vector success
changes by less than .02. The multi-event pool shows a larger task-blind shift,
from .240 to .623, while carrier vectors change from .755 to .773. These
outcomes demonstrate reward sensitivity without identifying a behavioral cause.

\subsection{Control beyond the baseline DFA input limit}
\cref{tab:xland-size} stratifies the multi-event test tasks by minimal-DFA
size. Fifteen tasks exceed the published RAD checkpoint's ten-state input
limit, while the Morphata operator retains 16 configuration slots. The four
largest-DFA tasks have high carrier-vector success under both rewards.
Task families and physical difficulty differ across strata, so these results
establish measured applicability rather than a monotone scaling law.

\begin{table}[htbp]
\centering\small\setlength{\tabcolsep}{4pt}
\caption{Multi-event XLand with 139 training tasks: success by minimal-DFA size, five policy
seeds and eight layouts per task. Parentheses give 95\% crossed-bootstrap
intervals. The largest stratum contains four tasks.}
\label{tab:xland-size}
\begin{tabular}{llccc}
\toprule
Reward & Conditioning & DFA $\leq10$ & DFA 11--32 & DFA 33--128\\
& Tasks & 17 & 11 & 4\\
\midrule
0 & Task-blind & 0.513 (0.294, 0.735) & 0.709 (0.463, 0.911) & 0.850 (0.425, 1.000)\\
0 & Ours/Initial & 0.547 (0.338, 0.750) & 0.745 (0.550, 0.908) & 0.694 (0.250, 0.950)\\
0 & Ours/Residual & 0.659 (0.439, 0.863) & 0.861 (0.710, 0.973) & 0.988 (0.900, 1.000)\\
0 & Ours/Carrier bits & 0.535 (0.322, 0.748) & 0.802 (0.623, 0.948) & 0.819 (0.275, 1.000)\\
0 & Ours/Carrier vectors & 0.651 (0.434, 0.855) & 0.880 (0.755, 0.969) & 0.994 (0.936, 1.000)\\
\addlinespace
-1 & Task-blind & 0.191 (0.000, 0.443) & 0.264 (0.000, 0.595) & 0.381 (0.000, 0.800)\\
-1 & Ours/Initial & 0.432 (0.218, 0.650) & 0.532 (0.295, 0.764) & 0.738 (0.267, 0.988)\\
-1 & Ours/Residual & 0.644 (0.422, 0.850) & 0.884 (0.756, 0.978) & 0.988 (0.900, 1.000)\\
-1 & Ours/Carrier bits & 0.503 (0.287, 0.716) & 0.768 (0.564, 0.927) & 0.819 (0.274, 1.000)\\
-1 & Ours/Carrier vectors & 0.629 (0.417, 0.834) & 0.870 (0.725, 0.978) & 0.969 (0.850, 1.000)\\
\addlinespace
\bottomrule
\end{tabular}

\end{table}

\subsection{Observation and network controls}
The separate view/encoder grid uses the RAD-compatible pool, five seeds,
1,000 updates, and the depth-one reference policy. \cref{tab:observation-controls}
includes the full-convolutional condition. These exploratory controls show no
uniform benefit from convolution or full observations. They differ in policy
architecture from the depth-three confirmation even though their frame budgets
match. Failed-learning seeds remain in the estimates.

\begin{table}[htbp]
\centering\small\setlength{\tabcolsep}{4pt}
\caption{XLand observation/network controls: mean test success and 95\%
crossed seed/task/layout intervals, five seeds, 17 tasks and eight layouts.
Local uses a $5\times5$ view; Full exposes the room. All conditions use
1,000 updates and rejection reward $-1$.}
\label{tab:observation-controls}
\begin{tabular}{llcc}
\toprule
View/encoder & Conditioning & 25 training tasks & 87 training tasks\\
\midrule
Local/linear & Task-blind & 0.313 (0.084, 0.587) & 0.100 (0.000, 0.327)\\
Local/linear & Ours/Residual & 0.631 (0.423, 0.825) & 0.647 (0.437, 0.843)\\
Local/linear & RAD/Residual & 0.671 (0.471, 0.850) & 0.724 (0.528, 0.893)\\
\addlinespace
Local/conv & Task-blind & 0.512 (0.306, 0.715) & 0.104 (0.000, 0.347)\\
Local/conv & Ours/Residual & 0.619 (0.413, 0.810) & 0.643 (0.432, 0.840)\\
Local/conv & RAD/Residual & 0.632 (0.425, 0.822) & 0.603 (0.384, 0.809)\\
\addlinespace
Full/linear & Task-blind & 0.004 (0.000, 0.026) & 0.000 (0.000, 0.000)\\
Full/linear & Ours/Residual & 0.435 (0.249, 0.626) & 0.462 (0.268, 0.663)\\
Full/linear & RAD/Residual & 0.475 (0.285, 0.672) & 0.413 (0.226, 0.609)\\
\addlinespace
Full/conv & Task-blind & 0.026 (0.000, 0.099) & 0.000 (0.000, 0.000)\\
Full/conv & Ours/Residual & 0.518 (0.319, 0.712) & 0.490 (0.291, 0.688)\\
Full/conv & RAD/Residual & 0.509 (0.307, 0.707) & 0.541 (0.340, 0.740)\\
\addlinespace
\bottomrule
\end{tabular}

\end{table}

\subsection{Corrected LetterWorld comparison}
LetterWorld uses RAD's task distribution, the large training pool, three
policy seeds, 28 test tasks, and eight fixed evaluation layouts. The sampler
rejects candidates until they satisfy its validity criterion; all evaluation
maps pass that criterion. Each update collects 128 steps from 256 environments;
611 updates give 20,021,248 transitions. This increases the rollout batch
relative to the 16-environment reference recipe. Validation runs every 15
updates with 32 episodes. The original network uses a linear observation
encoder and shared depth-one torso; the new network uses a convolutional
encoder, separate heads, depth-three actor, depth-two critic, and orthogonal
initialization. Both retain rejection reward $-1$ and a constant learning rate.

Recipe and new-network choices followed inspection of pilot test outcomes.
Environment count was compared on a one-seed, 5M-frame validation screen.
\cref{tab:letterworld-final,tab:letterworld-depth} report the
corrected comparison; the depth bins were chosen after inspecting test results.
The depth-three bin contains one task, and the deepest bin three, so their
intervals are particularly uncertain.

\begin{table}[htbp]
\centering\small\setlength{\tabcolsep}{4pt}
\caption{Corrected LetterWorld with the large training pool: test success with 95\% crossed
seed/task/layout intervals, three policy seeds, 28 tasks and eight layouts.
RAD-heavy s0/s1 are different encoder checkpoints; each has three policy seeds.
The uniform-random floor is .318 [.215, .430].}
\label{tab:letterworld-final}
\begin{tabular*}{\linewidth}{@{\extracolsep{\fill}}lcc@{}}
\toprule
Conditioning & Linear/shared & Conv/separate\\
\midrule
Task-blind & 0.290\,{\scriptsize (0.176--0.420)} & 0.308\,{\scriptsize (0.201--0.430)}\\
Ours/Residual & 0.693\,{\scriptsize (0.564--0.811)} & 0.702\,{\scriptsize (0.557--0.830)}\\
Ours/RAD-heavy s0 & 0.661\,{\scriptsize (0.518--0.795)} & 0.671\,{\scriptsize (0.527--0.808)}\\
Ours/RAD-heavy s1 & 0.671\,{\scriptsize (0.531--0.799)} & 0.671\,{\scriptsize (0.525--0.804)}\\
RAD/Residual & 0.876\,{\scriptsize (0.793--0.943)} & 0.872\,{\scriptsize (0.786--0.942)}\\
\bottomrule
\end{tabular*}

\end{table}
\begin{table}[htbp]
\centering\small\setlength{\tabcolsep}{4pt}
\caption{Post hoc LetterWorld depth breakdown for the new network. Distance
is the shortest event sequence to acceptance. Cells report mean success and
95\% crossed-bootstrap intervals; all 28 test tasks are included.}
\label{tab:letterworld-depth}
\begin{tabular*}{\linewidth}{@{\extracolsep{\fill}}lcccc@{}}
\toprule
Conditioning & Distance 2 & Distance 3 & Distance 4--5 & Distance $6+$\\
Tasks & 17 & 1 & 7 & 3\\
\midrule
Task-blind & 0.434\,{\scriptsize (0.294--0.588)} & 0.208\,{\scriptsize (0.000--0.500)} & 0.137\,{\scriptsize (0.028--0.268)} & 0.028\,{\scriptsize (0.000--0.150)}\\
Ours/Residual & 0.926\,{\scriptsize (0.861--0.979)} & 0.792\,{\scriptsize (0.000--0.958)} & 0.417\,{\scriptsize (0.142--0.708)} & 0.069\,{\scriptsize (0.000--0.250)}\\
Ours/RAD-heavy s0 & 0.904\,{\scriptsize (0.815--0.975)} & 0.333\,{\scriptsize (0.000--0.625)} & 0.417\,{\scriptsize (0.167--0.688)} & 0.056\,{\scriptsize (0.000--0.208)}\\
Ours/RAD-heavy s1 & 0.882\,{\scriptsize (0.780--0.964)} & 0.833\,{\scriptsize (0.000--1.000)} & 0.369\,{\scriptsize (0.129--0.646)} & 0.125\,{\scriptsize (0.000--0.375)}\\
RAD/Residual & 0.949\,{\scriptsize (0.887--0.991)} & 1.000\,{\scriptsize (0.000--1.000)} & 0.738\,{\scriptsize (0.483--0.931)} & 0.708\,{\scriptsize (0.000--1.000)}\\
\bottomrule
\end{tabular*}

\end{table}

RAD residuals exceed the original Morphata residual encoder by .170
[.068, .284] with the new network and .183 [.077, .298] with the original
network. The RAD-heavy encoders do not close this gap in the tested runs.
There is no paired test establishing equivalence between network conditions.
The new-network gap is .022 [-.042, .086] at distance two and .321
[.046, .589] at distances four to five; the deepest-bin estimate is .639
[.000, .875]. These are residual-embedding comparisons, not operator audits.

\subsection{Residual geometry and exploratory interventions}
The original ten-letter bank has 1,114 reachable rows, 724 nonzero rows used
in the geometry audit, and 675 languages. Among 14,196 different-language
pairs first distinguished by words of length 4--8, median standardized
Euclidean distance is .145 for Morphata and 4.87 for RAD. Distinguishing-word
length is distinct from shortest acceptance distance. Same-language distances
across presentations need not vanish, so exact learned canonicity is not
established. Raw and standardized geometry are kept separate.

The near-collision threshold (cosine distance below $10^{-3}$), depth bins,
and fine-tune interventions were chosen after inspecting test outcomes and
bank geometry, including test residuals. Starting from one 75k-step encoder,
three continued-training arms use discount .5, discount .9, or discount .9
with a distance-to-acceptance head (weight .5), each with a fresh optimizer.
\cref{tab:finetune} aligns geometry, policy outcomes, and fidelity at
15k additional steps. The 10k and 15k evaluations are checkpoints on the
same continued-training trajectories, not independent replications. The
baseline remains the encoder of record.

\begin{table}[htbp]
\centering\small\setlength{\tabcolsep}{4pt}
\caption{Exploratory ten-letter fine-tunes at 15k additional encoder steps:
one encoder seed, three policy seeds per arm. Success is corrected LetterWorld
with the new network; parentheses give crossed 95\% intervals. Fidelity
columns count audited rows/pairs and are not policy outcomes.}
\label{tab:finetune}
\begin{tabular}{lrrrr}
\toprule
Encoder & Near-collisions & Test success & \shortstack{Plain/product\\(/112,/71)} & \shortstack{Composition\\AND/OR (/30)}\\
\midrule
Baseline 75k & 3812 & .702 (.559--.830) & 104/66 & 29/24\\
Discount .5 & 4474 & .695 (.551--.826) & 107/66 & 29/23\\
Discount .9 & 174 & .686 (.540--.823) & 107/66 & 29/25\\
Discount .9 + progress & 123 & .749 (.616--.869) & 102/65 & 26/23\\
\bottomrule
\end{tabular}
\end{table}

Reducing near-collisions alone does not yield a resolved control improvement:
discount .9 minus baseline success is $-.016$ [$-.103,+.064$]. The progress
head gives $+.046$ [$-.065,+.156$], with fewer exactly equivalent plain and
product rows and conjunction compositions. At 10k steps the corresponding
successes were .683 and .763; all improvement intervals also included zero.
These interventions motivate further study;
they do not establish a repair or a causal account of the LetterWorld gap.

\subsection{Validation trajectories and seed variability}
\cref{fig:xland-zero,fig:xland-minus-one,fig:letterworld-final-curves} show the frozen
training-time validation exports. Bands are pointwise 95\% intervals from
2,000 seed resamples. They are distinct from the crossed test intervals.
\cref{tab:xland-seeds} records the range of final test means across all
five policy seeds for each condition.

\begin{table}[htbp]
\centering\small\setlength{\tabcolsep}{3.8pt}
\caption{Ranges of individual policy-seed test means for the final XLand
comparison. Each condition includes five seeds, including zero-success runs.
Large pools contain 87/139/173 tasks for RAD-compatible/multi-event/plain
tasks. These ranges are not confidence intervals.}
\label{tab:xland-seeds}
\begin{tabular}{llcccc}
\toprule
Pool & Conditioning & \shortstack{25 tasks\\Reward $0$} & \shortstack{25 tasks\\Reward $-1$} & \shortstack{Large pool\\Reward $0$} & \shortstack{Large pool\\Reward $-1$}\\
\midrule
RAD-compatible & Task-blind & 0.507--0.551 & 0.000--0.537 & 0.471--0.529 & 0.000--0.515\\
RAD-compatible & Ours/Initial & 0.507--0.544 & 0.456--0.522 & 0.515--0.581 & 0.360--0.478\\
RAD-compatible & Ours/Residual & 0.515--0.654 & 0.485--0.640 & 0.625--0.676 & 0.632--0.662\\
RAD-compatible & Ours/Carrier bits & 0.390--0.478 & 0.441--0.485 & 0.500--0.581 & 0.382--0.537\\
RAD-compatible & Ours/Carrier vectors & 0.544--0.654 & 0.529--0.625 & 0.640--0.662 & 0.596--0.669\\
RAD-compatible & RAD/Initial & 0.419--0.463 & 0.346--0.419 & 0.456--0.566 & 0.397--0.471\\
RAD-compatible & RAD/Residual & 0.581--0.625 & 0.537--0.676 & 0.684--0.735 & 0.676--0.735\\
\addlinespace
Multi-event & Task-blind & 0.574--0.645 & 0.555--0.598 & 0.590--0.648 & 0.000--0.625\\
Multi-event & Ours/Initial & 0.535--0.621 & 0.379--0.504 & 0.582--0.656 & 0.453--0.539\\
Multi-event & Ours/Residual & 0.629--0.730 & 0.641--0.719 & 0.746--0.797 & 0.738--0.785\\
Multi-event & Ours/Carrier bits & 0.473--0.520 & 0.453--0.547 & 0.633--0.703 & 0.598--0.680\\
Multi-event & Ours/Carrier vectors & 0.629--0.691 & 0.637--0.684 & 0.738--0.793 & 0.711--0.773\\
\addlinespace
Plain & Task-blind & 0.720--0.764 & 0.733--0.764 & 0.716--0.770 & 0.743--0.770\\
Plain & Ours/Initial & 0.709--0.780 & 0.689--0.770 & 0.757--0.787 & 0.764--0.784\\
Plain & Ours/Residual & 0.706--0.784 & 0.713--0.777 & 0.811--0.841 & 0.811--0.841\\
Plain & Ours/Carrier bits & 0.662--0.747 & 0.615--0.723 & 0.753--0.764 & 0.740--0.770\\
Plain & Ours/Carrier vectors & 0.625--0.689 & 0.588--0.682 & 0.814--0.851 & 0.814--0.851\\
\addlinespace
\bottomrule
\end{tabular}

\end{table}

\ifextendedpaper\else
\begin{figure}[!t]
\centering%
  \includestandalone[]{figures/standalone/xland-final-zero}
\caption{XLand validation trajectories, rejection reward 0. Rows show task
pools; columns show 25 training tasks (left) and 87/139/173 training tasks
(right, for RAD-compatible/multi-event/plain). All five policy seeds are included. Bands are pointwise 95\% intervals
from 2,000 policy-seed bootstrap resamples. The selected architecture is linear/shared depth three.}
\label{fig:xland-zero}
\end{figure}

\fi
\begin{figure}[p]
\centering%
  \includestandalone[]{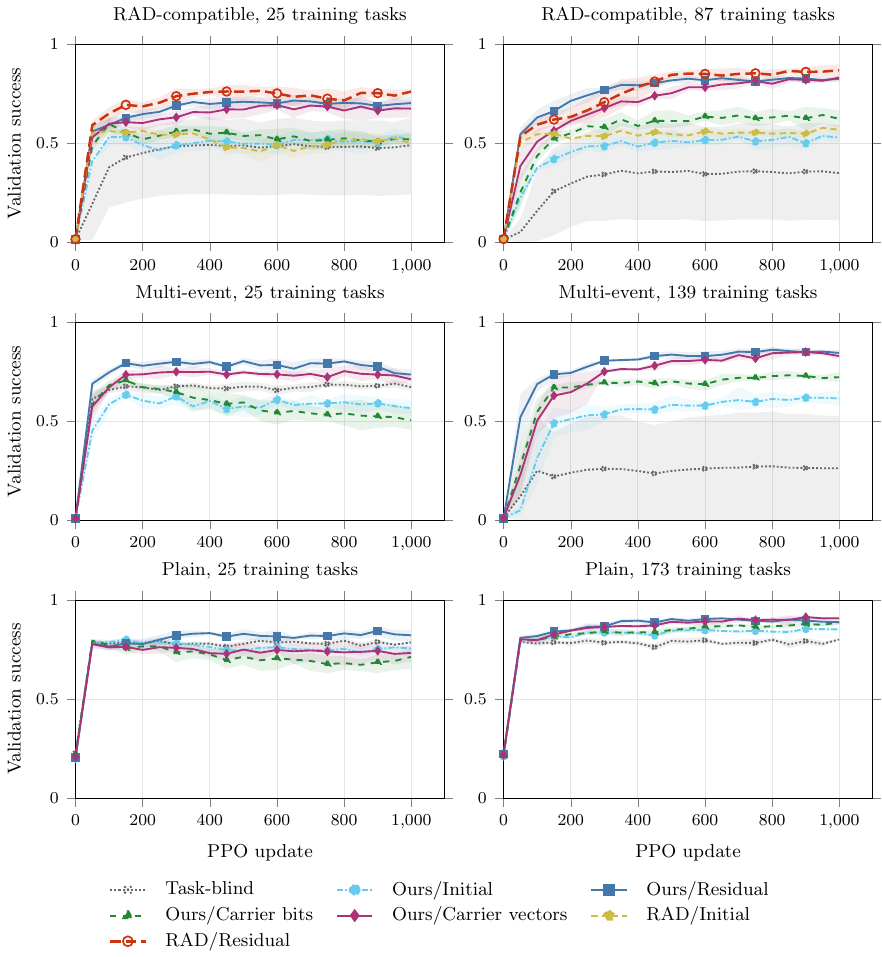}
\caption{XLand validation trajectories, reference rejection reward $-1$,
under the same pool, seed, and budget protocol as \cref{fig:xland-zero}.}
\label{fig:xland-minus-one}
\end{figure}
\begin{figure}[p]
\centering%
  \begin{tabular*}{\linewidth}{@{\extracolsep{\fill}}lcc@{}}
\toprule
Conditioning & Linear/shared & Conv/separate\\
\midrule
Task-blind & 0.290\,{\scriptsize (0.176--0.420)} & 0.308\,{\scriptsize (0.201--0.430)}\\
Ours/Residual & 0.693\,{\scriptsize (0.564--0.811)} & 0.702\,{\scriptsize (0.557--0.830)}\\
Ours/RAD-heavy s0 & 0.661\,{\scriptsize (0.518--0.795)} & 0.671\,{\scriptsize (0.527--0.808)}\\
Ours/RAD-heavy s1 & 0.671\,{\scriptsize (0.531--0.799)} & 0.671\,{\scriptsize (0.525--0.804)}\\
RAD/Residual & 0.876\,{\scriptsize (0.793--0.943)} & 0.872\,{\scriptsize (0.786--0.942)}\\
\bottomrule
\end{tabular*}

\caption{Corrected LetterWorld validation trajectories, three policy seeds.
Panels distinguish network and encoder-bank conditions; baseline encoder
uses \texttt{c3lsem10-s0}, and 10r-s0/s1 are RAD-heavy checkpoints.
Bands resample policy seeds only.}
\label{fig:letterworld-final-curves}
\end{figure}

\begin{figure}[t]
\centering
\includegraphics[width=\linewidth,height=.60\textheight,keepaspectratio]{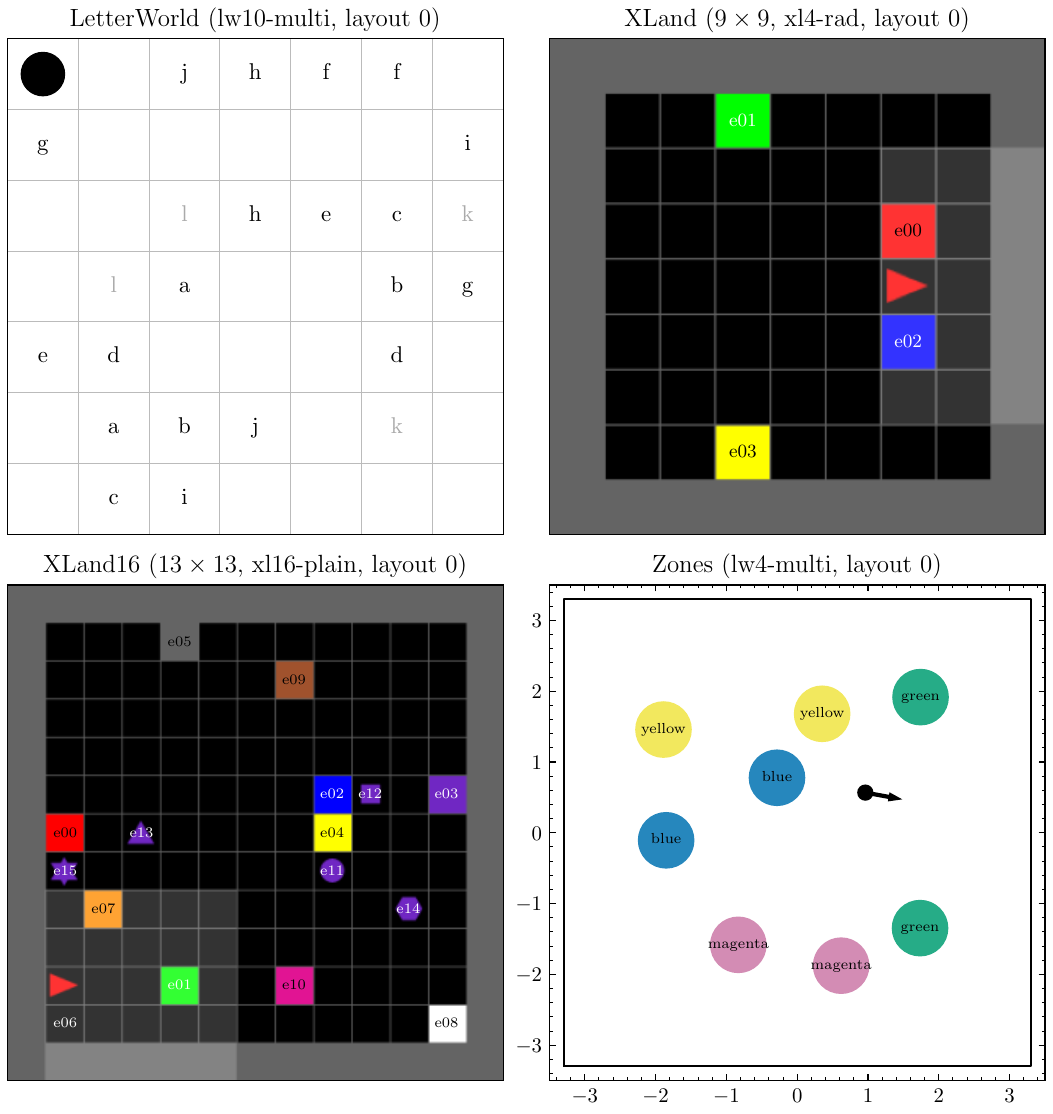}
\caption{Representative layouts: LetterWorld and four-letter XLand (top),
sixteen-letter XLand and Zones (bottom). Objects, letters, and zones generate
monitor events. The quantitative control results here use four-letter XLand
and ten-letter LetterWorld.}
\label{fig:rl-environments}
\end{figure}

\fi

\end{document}